\documentclass[11pt]{article}
\usepackage{graphicx}
\usepackage{epsfig}
\usepackage{amsfonts}
\usepackage{amssymb}
\usepackage{amsmath,amssymb}
\usepackage{url}
\usepackage{graphicx,epsf, epsfig, amssymb, amsmath, amsfonts, mathrsfs}
\usepackage{longtable}
\usepackage{bm}
\usepackage{color}
\usepackage{enumitem}
\usepackage{float}
\usepackage{caption}
\usepackage{subcaption}
\usepackage{upgreek}
\usepackage{cite}
\usepackage{mathtools}
\usepackage{titlesec}
\usepackage{lipsum}
\usepackage{xcolor}
\usepackage[section]{placeins}
\usepackage[colorlinks=true,linktocpage=true,linkcolor=blue,citecolor=blue,urlcolor=blue]{hyperref}
\usepackage{tensor}
\usepackage{multirow}
\usepackage[utf8]{inputenc}
\usepackage{authblk}
\usepackage{array}
\usepackage{breqn}

\newcommand{\beq}{\begin{equation}}
	\newcommand{\eeq}{\end{equation}}
\newcommand{\beqa}{\begin{eqnarray}}
	\newcommand{\eeqa}{\end{eqnarray}}

\newcommand{\dmoff}[1]{}

\usepackage{comment}

\begin{document}

\title{\Large{\bf 
Two asymptotically flat spinning black holes\\
balanced by their self-interacting, synchronised scalar hair
} }

\author{
{\large Chen Liang}$^{1}$,
{\large Carlos Herdeiro}$^{1,2}$  and  
{\large Eugen Radu}$^{1}$
\vspace*{0.2cm}
\\
$^{1}${\small Departamento de Matem\'atica da Universidade de Aveiro and } 
\\ {\small  Centre for Research and Development  in Mathematics and Applications (CIDMA),} 
\\ {\small    Campus de Santiago, 3810-193 Aveiro, Portugal}
\\ $^{2}${\small Programa de P\'os-Gradua\c{c}\~{a}o em F\'{\i}sica, Universidade 
	Federal do Par\'a, 66075-110, Bel\'em, Par\'a, Brazil}
}

\date{May 2026}

\maketitle

\begin{abstract}
Asymptotically flat \textit{balanced} configurations of two spinning black holes with synchronised scalar hair (2sBHs) are possible~\cite{Herdeiro:2023roz}. These are constructed within a generalized Bach-Weyl framework and arise from two spinning boson stars (2sBSs) by placing a horizon at the center of each component. Here, we investigate the effects of quartic scalar self-interactions on this family of solutions, comprising the 2sBSs, the 2sBHs, and an intermediate configuration---single spinning black hole with quadrupolar scalar hair (1sBHs). For 2sBSs, the additional repulsive force introduced by the self-interactions drives a topological transition of the ergoregion, from a single torus to a double torus, in the strong-gravity regime. For 1sBHs, as the self-interaction coupling strength increases, the solutions become ``hairier'' but their horizons cannot become heavier; moreover, the self-interactions broaden the regime in which an analytical effective model accurately describes these solutions~\cite{Herdeiro:2017phl, Brihaye:2018woc}. For 2sBHs, increasing the coupling reshapes the bifurcation structure of the solution sequences and, as in the 1sBH case, repulsive self-interactions cannot make the horizons heavier; horizons carrying a larger mass fraction are obtained only when attractive self-interactions are considered.
  \end{abstract}

\tableofcontents

 \section{Introduction}
Scalar fields have played a central role in general relativity (GR) over the past decades. Part of this interest comes from high-energy physics and cosmology: string theory compactifications predict a large number of ultralight bosonic fields~\cite{Arvanitaki:2009fg,Svrcek:2006yi}, axion-like particles are widely studied as dark matter candidates~\cite{Marsh:2015xka,Ferreira:2020fam}, and ultralight scalars also underlie the fuzzy dark matter scenario on galactic scales~\cite{Hu:2000ke,Hui:2016ltb,Hui:2021tkt}. These motivations make the behavior of scalar fields in strong-gravity regimes an active research direction. In this context, scalar fields offer simple counterexamples to the no-hair theorems~\cite{Bekenstein:1995un,Sotiriou:2015pka,Herdeiro:2015waa,Cardoso:2016ryw} and reveal equilibrium configurations beyond the Kerr family~\cite{Chrusciel:2012jk,Herdeiro:2014goa,Herdeiro:2015gia,Volkov:2016ehx,Burrage:2023zvk}. In addition, the superradiant instability of rotating black holes (BHs)~\cite{Hod:2012px,Benone:2014ssa,Brito:2015oca} provides a natural mechanism to produce bound scalar configurations around BHs. With the development of gravitational-wave astronomy~\cite{LIGOScientific:2016aoc,LIGOScientific:2021rnv} and horizon-scale imaging by 
the Event Horizon Telescope~\cite{EventHorizonTelescope:2019dse,EventHorizonTelescope:2022wkp}, it has also become relevant to ask whether ultracompact bosonic objects can serve as BH mimickers~\cite{Liebling:2012fv,Cunha:2017wao,Olivares:2018abq,Cardoso:2019rvt,Jaramillo:2026ygy}. These developments together
motivate systematic studies of boson stars (BSs), scalar hair, and the associated BH spacetimes.

Among scalar field solutions in GR, BSs are one of the most studied examples. They are self-gravitating solitons of a 
massive complex scalar field, first constructed in the spherically symmetric case~\cite{Kaup:1968zz, Ruffini:1969qy} and later extended to the spinning case~\cite{Schunck:1996he, Yoshida:1997qf}. A major step forward was the construction of Kerr BHs with 
synchronised scalar hair~\cite{Herdeiro:2014goa}, obtained by placing an event horizon at the center of a spinning BS under the synchronisation condition $\omega = m\Omega_H$, where $\omega$ is the scalar frequency, $m$ is the azimuthal harmonic index, and $\Omega_H$ is the horizon angular velocity. This condition also marks the threshold of the superradiant instability of Kerr BHs against massive scalar perturbations~\cite{Hod:2012px}, so these hairy BHs can be interpreted as nonlinear endpoints of superradiance. Within the Einstein-Klein-Gordon framework, placing a horizon inside a BS has since become a systematic way to construct hairy BHs associated with a given BS family, and this strategy has been applied to excited-state BSs~\cite{Wang:2018xhw}, gauged BSs~\cite{Herdeiro:2020xmb}, and self-interacting BSs~\cite{Herdeiro:2015tia}. Hairy BHs have also been constructed in various extensions beyond the Einstein-Klein-Gordon framework~\cite{Herdeiro:2016tmi, Gervalle:2024yxj, Herdeiro:2018daq, Kunz:2019sgn}.

\medskip

A particularly interesting recent member of the BS family consists of two spinning BSs (2sBSs) placed along a common axis with opposite phases. In our previous work~\cite{Herdeiro:2023roz}, starting from 2sBSs, a family of two spinning black holes (2sBHs) in equilibrium was constructed, whose horizons emerge at equilibrium points of the scalar environment provided by the 2sBS. This is a nontrivial result because, within vacuum GR, asymptotically flat stationary 2sBHs are known to necessarily contain a conical singularity on the axis between the two BHs, representing an unphysical strut needed 
to balance the mutual gravitational attraction~\cite{Dietz,Costa:2009wj,Hennig:2011fp,Neugebauer:2011qb}. Several strategies have been proposed to remove this singularity, such as introducing an external gravitational field that provides an additional 
force on the BHs~\cite{Astorino:2021dju}, or considering non-asymptotically-flat backgrounds and/or gauge charges~\cite{Majumdar:1947eu,Papapetrou:1948jw,Dias:2023rde,Biggs:2024yqf}. The 2sBS scenario offers a different route, in which the scalar environment --- which is itself compact  --  balances the 2sBHs in a regular and asymptotically flat way.

A further natural direction is to consider scalar self-interactions, motivated from both astrophysical and theoretical perspectives. The maximal ADM mass of mini-BSs---BSs made up of a free complex scalar field---is of the order of $M_{\rm Pl}^2/\mu$, which is too small for astrophysical relevance unless $\mu$ is extremely light. As first pointed out by Colpi et al~\cite{Colpi:1986ye}, adding a repulsive quartic self-interaction of the form $\lambda|\Psi|^4$ raises this upper bound to $\sqrt{\lambda}\,M_{\rm Pl}^3/\mu^2$, so that BSs with Standard-Model-range scalar masses can reach astrophysically interesting values. Additionally, recent nonlinear dynamical evolutions have shown that self-interactions can mitigate dynamical instabilities of various BS models, including spherical excited BSs~\cite{Sanchis-Gual:2021phr, Brito:2023fwr}, single spinning BSs~\cite{Siemonsen:2020hcg}, and dipolar BSs~\cite{Ildefonso:2023qty}. The effect of quartic self-interactions on spinning BSs and the associated hairy Kerr BHs has been studied in~\cite{Herdeiro:2015tia}, where it was found that self-interactions make the scalar matter distribution more 
extended, and reshape the domain of existence of the hairy Kerr BHs in a nontrivial way---with the total ADM mass growing with $\lambda$ but the horizon mass not following this growth, so that such BHs become ``hairier but not heavier''. In contrast, the effects of scalar self-interactions on the hairy BHs associated with 2sBSs have not yet been explored. This gap is nontrivial, because 2sBSs differ qualitatively from standard spinning BSs in both geometry and topology: they consist of two rotating lumps of scalar matter with opposite phases arranged along a common axis, a structure that supports not only single spinning BHs (1sBHs) with quadrupolar scalar hair~\cite{Kunz:2019bhm} but also 2sBHs in equilibrium~\cite{Herdeiro:2023roz}. Furthermore, given the rich effects of self-interactions already observed for 2sBSs themselves~\cite{Sun:2025eis}, one may expect even more pronounced consequences once horizons are introduced.

\medskip

One of the goals of this paper is to investigate how quartic self-interactions affect 2sBSs. The additional repulsive force they introduce modifies the scalar matter distribution and, in particular, changes the topology of the ergoregion. Then, we also consider how quartic self-interactions affect the hairy BHs associated with 2sBSs. The results in~\cite{Herdeiro:2015tia} showed that for hairy BHs within spinning BSs, such self-interactions can significantly increase the total mass of hairy BHs, but they do not enlarge the horizon mass and therefore do not make the BHs heavier. Motivated by this, our second goal is to examine the effects of self-interactions on 1sBHs and 2sBHs with quadrupolar scalar hair.

\medskip

The paper is organized as follows. Section~\ref{sec:framework} presents the Einstein-Klein-Gordon model with quartic self-interactions. In Section~\ref{sec:2sBS}, we construct the 2sBSs and analyze how the self-interactions modify their physical properties. Section~\ref{sec:1sBHs} presents the domain of existence of the 1sBHs with quadrupolar scalar hair, examines the effects on the horizon, and discusses how the strength of the self-interactions affect the accuracy of an effective model. In Section~\ref{sec:2sBHs}, we describe the numerical strategy used to construct the 2sBHs, and show how the self-interactions alter the bifurcation structure of the solution sequences as well as their physical properties. We close in Section~\ref{conclusions} with a summary of our main findings and a discussion of several open questions.

\section{Framework}\label{sec:framework}
In order to construct the hairy BHs, we consider a complex scalar field $\Psi$ minimally coupled to the Einstein-Hilbert action:
\begin{equation}
    S = \int d^4x \sqrt{-g}\left( \frac{R}{16 \pi G} - g^{\mu\nu}\Psi^*_{,\mu}\Psi_{,\nu} - U(|\Psi|^2) \right)\,,
\end{equation}
where $R$ is the Ricci scalar and $G$ is the gravitational constant. The scalar field potential $U(|\Psi|^2)$ includes the mass term and self-interactions of the field. The equations of motion for the system are given by
\begin{subequations}\label{equek}
\begin{align}
\label{eque}
&
R_{\mu\nu} - \frac{1}{2}g_{\mu\nu}R-8\pi G T_{\mu\nu}=0\,, \\
\label{equk}
&
\left(\nabla^2 - \frac{dU}{d|\Psi|^2} \right)\Psi = 0\,, 
\end{align}
\end{subequations}
where 
\begin{eqnarray} \label{energy-momentum}
T_{\mu\nu} = \Psi^*_{,\mu}\Psi_{,\nu} + \Psi^*_{,\nu}\Psi_{,\mu} - g_{\mu\nu}\left[ \frac{1}{2}g^{\alpha\beta} \left( \Psi^*_{,\alpha}\Psi_{,\beta} + \Psi^*_{,\beta}\Psi_{,\alpha} \right) + U(|\Psi|^2) \right]\,.
\end{eqnarray} 
In this work, we consider a scalar field potential of the following form:
\begin{eqnarray} \label{potential}
U(|\Psi|^2)= \mu^2|\Psi|^2 + \lambda|\Psi|^4\,, 
\end{eqnarray} 
where $\mu$ is the mass of the scalar field and $\lambda > 0$ is the coupling constant that determines the strength of the repulsive self-interaction.

Under a global U(1) transformation, $\Psi \to \Psi e^{i\alpha}$, where $\alpha$ is an arbitrary constant, the action remains invariant. This symmetry yields the conserved current
\begin{equation}
j^\mu = \frac{i}{2}\left(\Psi^* \partial^\mu \Psi - \Psi \partial^\mu \Psi^*\right)\,.
\end{equation}
The Noether charge can be obtained by integrating the timelike component of this current over a spacelike hypersurface $\Sigma$:
\begin{equation}
Q = \int_\Sigma j^t \sqrt{-g} \, d^3x\,.
\end{equation}

In the following sections, based on the above model, we construct in turn the 2sBSs; the 1sBHs, obtained by placing a horizon at the center of a 2sBS; and the 2sBHs, obtained by placing two aligned horizons along the symmetry axis of a 2sBS. The differences between these configurations require different numerical approaches to construct them, which we describe in detail in each section, with particular emphasis on the more involved case of the 2sBHs, while discussing the impact of the self-interaction on each configuration.

\section{Two spinning boson stars}\label{sec:2sBS}
\subsection{Ansatz and boundary conditions}
Before constructing hairy BHs, we first consider the horizonless case. We use a metric ansatz that possesses two commuting Killing vector fields $\xi = \partial_t$ and $\eta = \partial_\varphi$~\cite{Herdeiro:2015gia}:
\begin{eqnarray}\label{metric}
ds^2= -e^{2F_0} dt^2 + 
e^{2F_1}\left(dr^2+r^2d\theta^2 \right)+
e^{2F_2}r^2\sin^2\theta\left(d\varphi-Wdt\right)^2
\,,
\end{eqnarray}
where the functions $F_0,F_1,F_2,W$ depend only on $r$ and $\theta$. The ansatz for the scalar field is
\begin{eqnarray}\label{matter_ansatz}
\Psi=\psi(r,\theta)e^{i(m\varphi-\omega t)}
\,,
\end{eqnarray}
where $m\in \mathbb{Z}$ is the azimuthal winding number and $\omega$ is the frequency. In the present study, we set $m=1$.

We focus on 2sBSs, configurations consisting of two scalar components arranged along a common axis. Taking into account the symmetry of the configuration and the
requirement of regularity, the equations~\eqref{equek} are solved subject to the following boundary conditions:
\begin{itemize}
  \item At the origin, the boundary conditions are compatible with regularity:
   \begin{align}
\partial_r F_0|_{r=0}=\partial_r F_1|_{r=0}=\partial_r F_2|_{r=0}=W|_{r=0}=0\,,\quad \psi|_{r=0}=0
\,.
\end{align}
  \item Asymptotic flatness requires that all functions vanish at infinity:
\begin{align}
F_0|_{r\to \infty}=F_1|_{r\to \infty}=F_2|_{r\to \infty}=W|_{r\to \infty}=0\,,\quad \psi|_{r\to \infty}=0
\,.
\end{align}
   \item On the symmetry axis, regularity and axial symmetry require
\begin{align}
\partial_\theta F_0|_{\theta=0,\pi}=\partial_\theta F_1|_{\theta=0,\pi}=\partial_\theta F_2|_{\theta=0,\pi}=\partial_\theta W|_{\theta=0,\pi}=0\,,\quad \psi|_{\theta=0,\pi}=0
\,.
\end{align}
Additionally, the constraint $F_1|_{\theta=0,\pi}=F_2|_{\theta=0,\pi}$ is imposed to avoid conical singularities.

  \item All solutions discussed in this work possess $\mathbb{Z}_2$ symmetry under reflections with respect to the equatorial plane; thus we impose the following conditions at $\theta=\pi/2$:
\begin{align}
\partial_\theta F_0|_{\theta=\frac{\pi}{2}}=\partial_\theta F_1|_{\theta=\frac{\pi}{2}}=\partial_\theta F_2|_{\theta=\frac{\pi}{2}}=\partial_\theta W|_{\theta=\frac{\pi}{2}}=0
\,,
\end{align}
and the quadrupolar scalar hair requires $\psi|_{\theta=\frac{\pi}{2}}=0$.
\end{itemize}
\subsection{Physical quantities}
The ADM mass $M$ and angular momentum $J$ can be read off from the asymptotic behaviour of the metric functions $g_{tt}$ and $g_{t\varphi}$:
\begin{align}\label{asymptotic}
g_{tt}=-1 + \frac{2GM}{r}+\ldots \,,\qquad g_{t\varphi}=-\frac{2GJ}{r}\sin^2\theta +\ldots
\,.
\end{align}
These quantities can also be calculated from the Komar integrals:
\begin{align}
M=\int \left( T_\mu^\mu - 2T_t^t \right)\sqrt{-g}\,dr\, d\theta\, d\phi
\,,\qquad
J=\int T_\phi^t \sqrt{-g}\,dr\, d\theta\, d\phi
\,.
\end{align}
The angular momentum and Noether charge obey the relation $J=mQ$.

Moreover, we define a \textit{distance} $D$ as the proper distance between the centers of the two components of a 2sBS, identified as the points where the symmetry axis intersects the planes at which the energy density $-T_t^t$ reaches its maximum value:
\begin{align}
D=2\int_{0}^{z_{\mathrm{max}}} \sqrt{g_{rr}(r,0)}\,dr
\,,
\end{align}
where the factor of $2$ arises from the $\mathbb{Z}_2$ symmetry of the system.
\subsection{Numerical results}

By appropriately rescaling the variables and parameters of the model, one can reduce the number of free parameters. It is convenient to introduce the following rescalings:
\begin{align}\label{rescalings}
\psi = \frac{\bar{\psi}}{\sqrt{4\pi G}} \,,\quad \lambda = 4\pi G \mu^2 \bar{\lambda}\,,\quad \omega = \bar{\omega}\mu \,,\quad r= \frac{\bar{r}}{\mu}
\,,
\end{align}
which amounts to setting $4\pi G = \mu = 1$ in the equations of motion. In what follows, we work with the rescaled quantities and drop the bars for notational simplicity.

After substituting the ansatz introduced in the previous section and appropriately combining the components of the Einstein equations~\eqref{eque}, one obtains a set of partial differential equations in which each equation contains second derivatives with respect to only one unknown function. We solve them numerically together with the matter equation~\eqref{equk}. 
In addition, \eqref{eque} implies the existence of two contraint equations which are not solved directly, being used to monitor the accuracy of the numerical results \cite{Herdeiro:2015gia}.
All solutions in this work were obtained using a professional finite difference solver
~\cite{Schonauer:1989zwe,SCHONAUER1990279,SCHONAUER2001473},  
which employs a Newton-Raphson method.
We introduce a compactified radial coordinate
\begin{align}
x=\frac{r}{r+c}
\,,
\end{align}
where $c$ is an arbitrary constant. This coordinate maps the semi-infinite region $[0,\infty)$ to the finite interval $[0,1]$. The equations of motion are discretised on a grid in $x$ and $\theta$. We use typical grids with sizes around $250 \times 50$ points. The relative error for the solutions reported in this work is less than $10^{-3}$.

\begin{figure}[h!]
  \centering
  \includegraphics[width=0.49\textwidth]{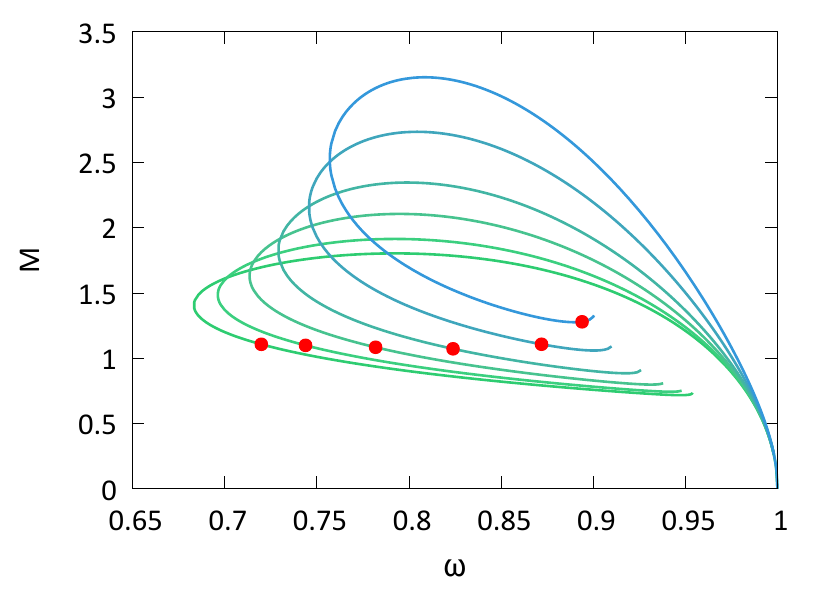}
  \includegraphics[width=0.49\textwidth]{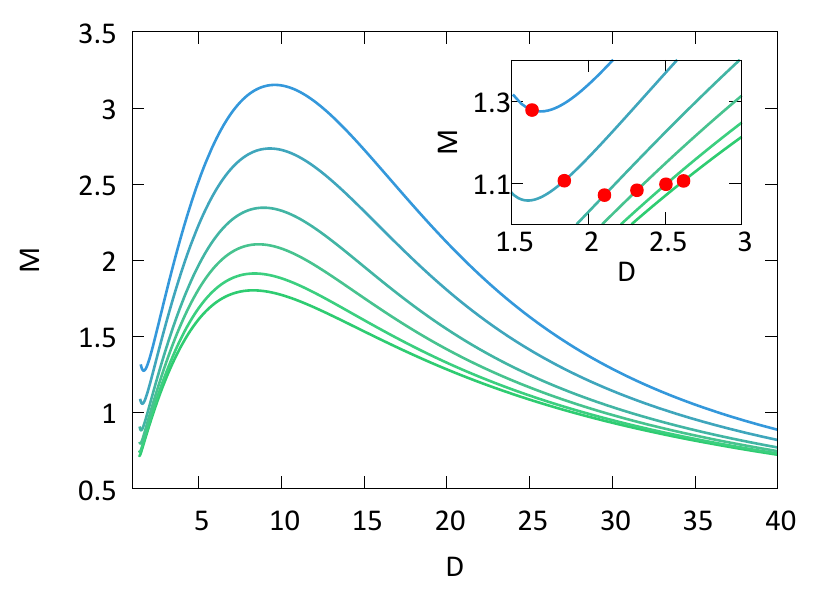}
  \caption{ADM mass  $vs.$ frequency (left panel) and proper distance $vs.$ ADM mass (right panel) for 2sBSs with $\lambda=0,5,15,30,60,100$, from bottom to top. The onset of ergoregions is indicated by the red points. The inset shows that ergoregions appear only at very small values of $D$.
  }
  \label{wMMD}
\end{figure}

\begin{figure}[h!]
  \centering
  \vspace{0.02\textheight}
  \includegraphics[height=0.3\textwidth]{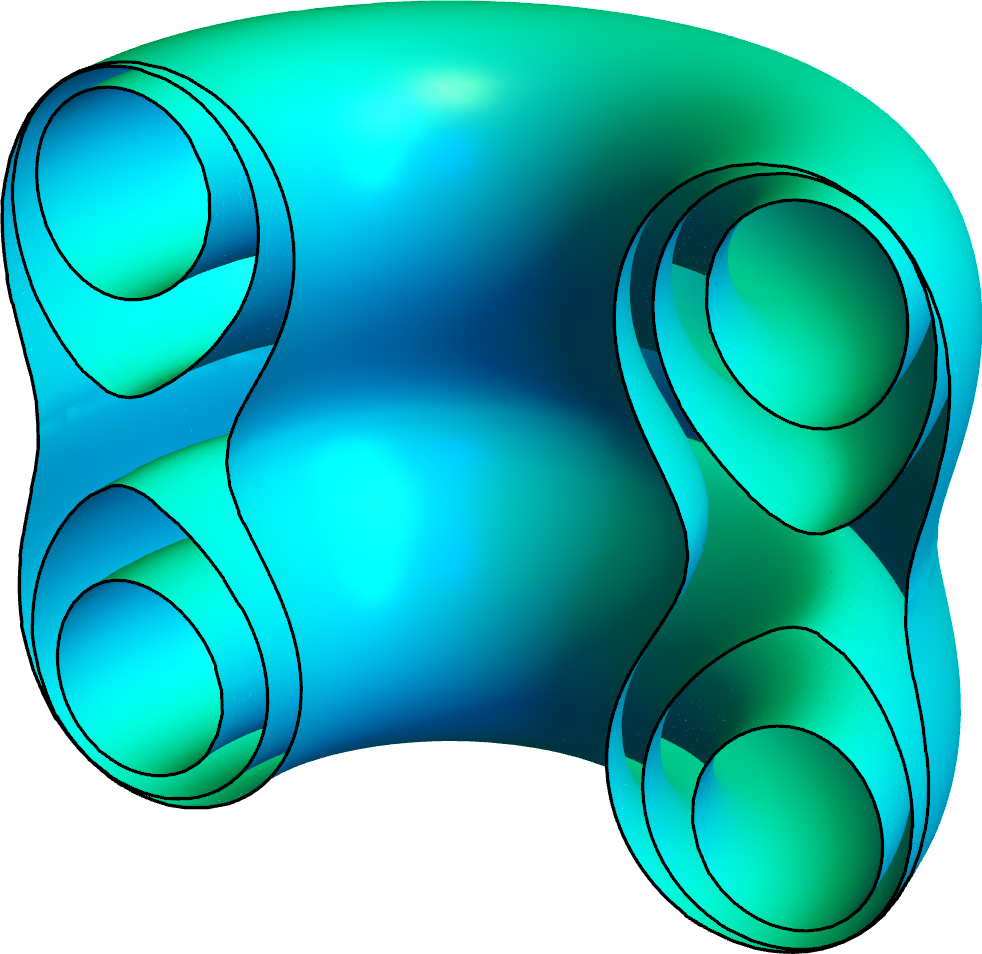}
  \caption{Ergosurfaces of 2sBSs with $\omega=0.78$ and $\lambda=0,5,10$, from the outermost to the innermost surface.}
  \label{ergo}
\end{figure}

In Fig.~\ref{wMMD}, we show the ADM mass and the proper distance between two components of 2sBSs. Similar to single spinning BSs~\cite{Kleihaus:2007vk}, due to an additional repulsive force caused by the self-interactions, the maximum mass of 2sBSs increases with the coupling $\lambda$. Moreover, for different values of $\lambda$, the distance is always around $D=10$ at maximum mass. In strong gravity regions, which correspond to the solution branches on the left side of the mass peaks in the right panel, the distance approaches zero. As $D$ increases, the mass decreases and the system approaches the Newtonian limit. For a given $D$, the ADM mass always increases with the coupling. This means that the quartic self-interactions not only provide a repulsion to balance the gravitational attraction of a single spinning BS, but can also enhance the repulsion between the two components of a 2sBS.

Additionally, the data points corresponding to 
the red dots
in the figure indicate 
the onset
of ergoregions that arise in strong gravity regions. The boundary of an ergoregion, or ergosurface, is defined by
\begin{align}
g_{tt}= -e^{2F_0} + e^{2F_2}W^2 r^2 \sin^2\theta = 0
\,.
\end{align}
Fig.~\ref{ergo} shows the ergosurfaces of 2sBSs. For a given $\omega$, the ergosurface transitions from a single torus to a double torus as the coupling increases. This topological change originates from the repulsive nature of the self-interactions, which suppress the accumulation of scalar matter in the region between the two components and thereby modify the local spacetime geometry such that the surface $g_{tt}=0$ disconnects along the symmetry axis.

Following the effective potential method of~\cite{Cunha:2016bjh}, we have also examined the light rings (LRs) of 2sBSs and confirmed the results of~\cite{Huang:2024gtu}: 2sBSs admit at most two stable LRs, symmetric under reflection across the equatorial plane, and two unstable LRs lying on the equatorial plane. A detailed study of how self-interactions affect the LR structure---in particular, in connection with the topological transition of the ergoregion discussed above---is left for future work.

\begin{figure}[h!]
  \centering
  \includegraphics[width=0.31\textwidth]{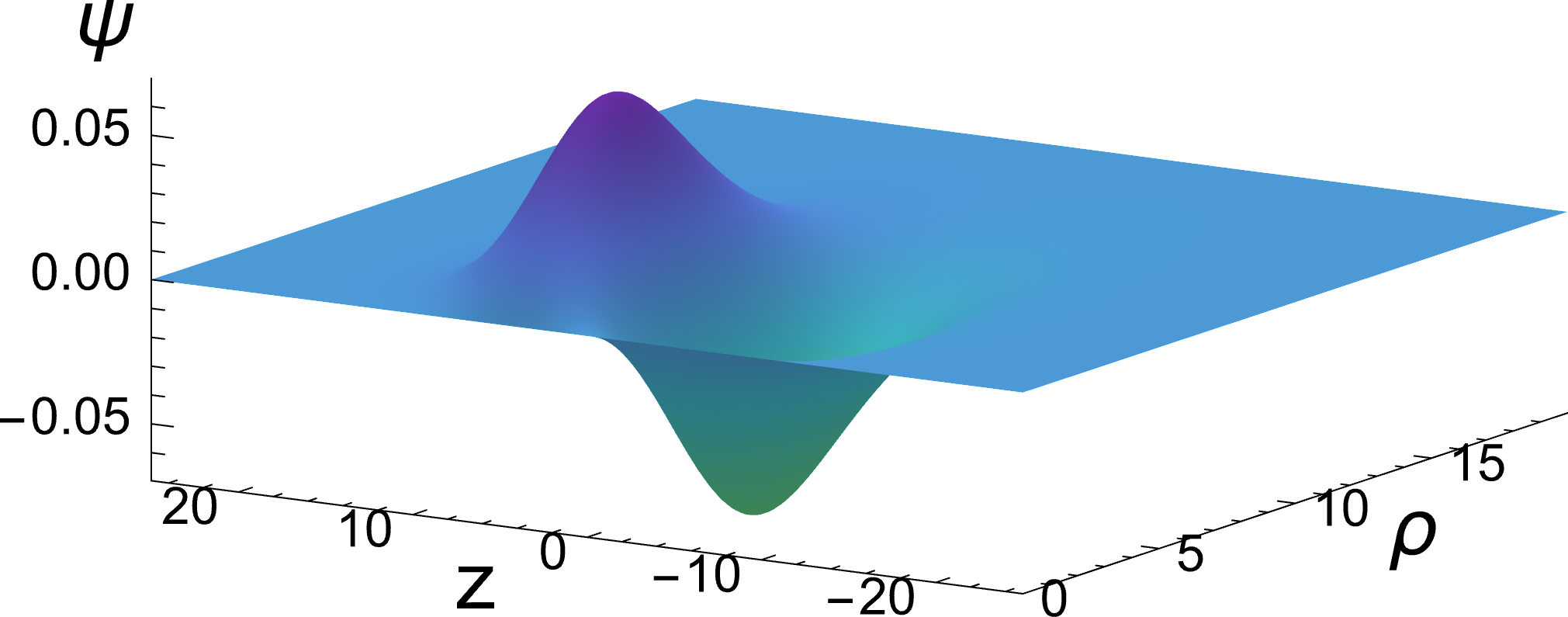}
    \hspace{0.01\textwidth}
  \includegraphics[width=0.31\textwidth]{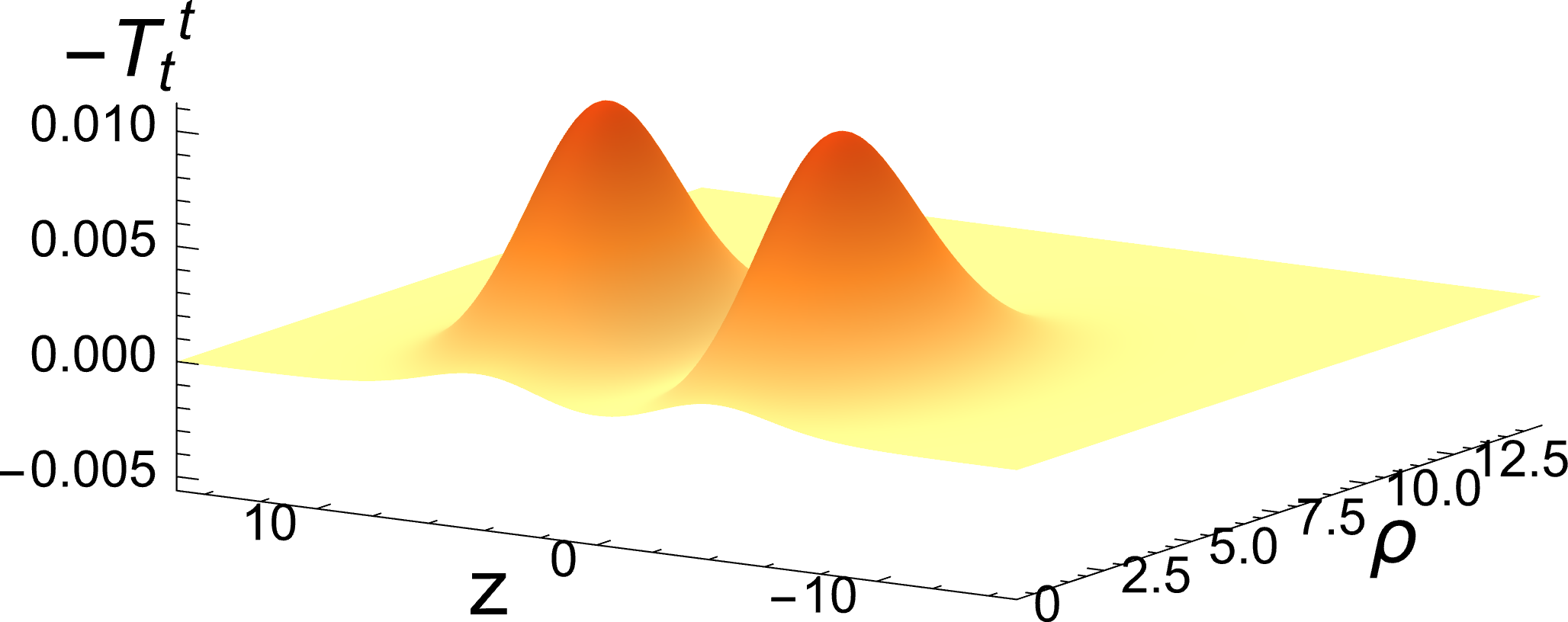}
    \hspace{0.01\textwidth}
  \includegraphics[width=0.31\textwidth]{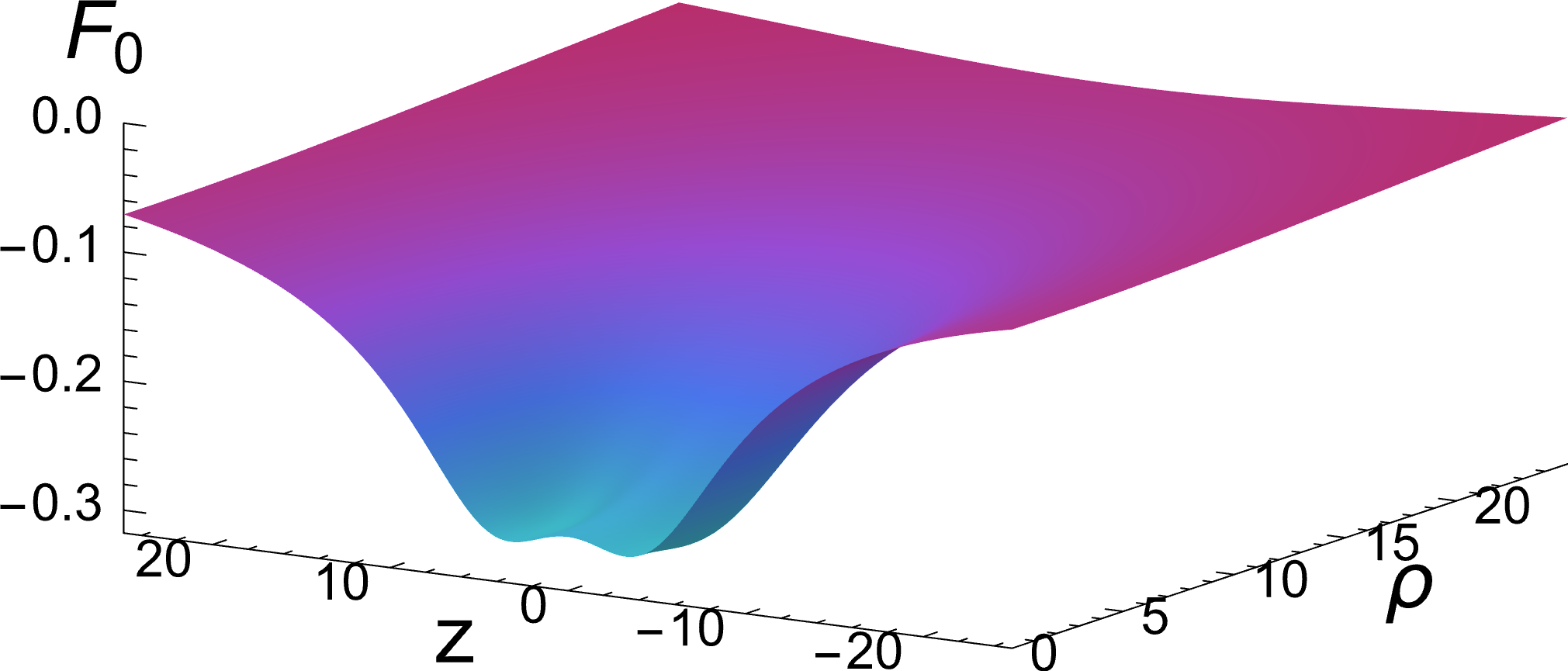}\\
  \vspace{0.05\textwidth}
  \includegraphics[width=0.31\textwidth]{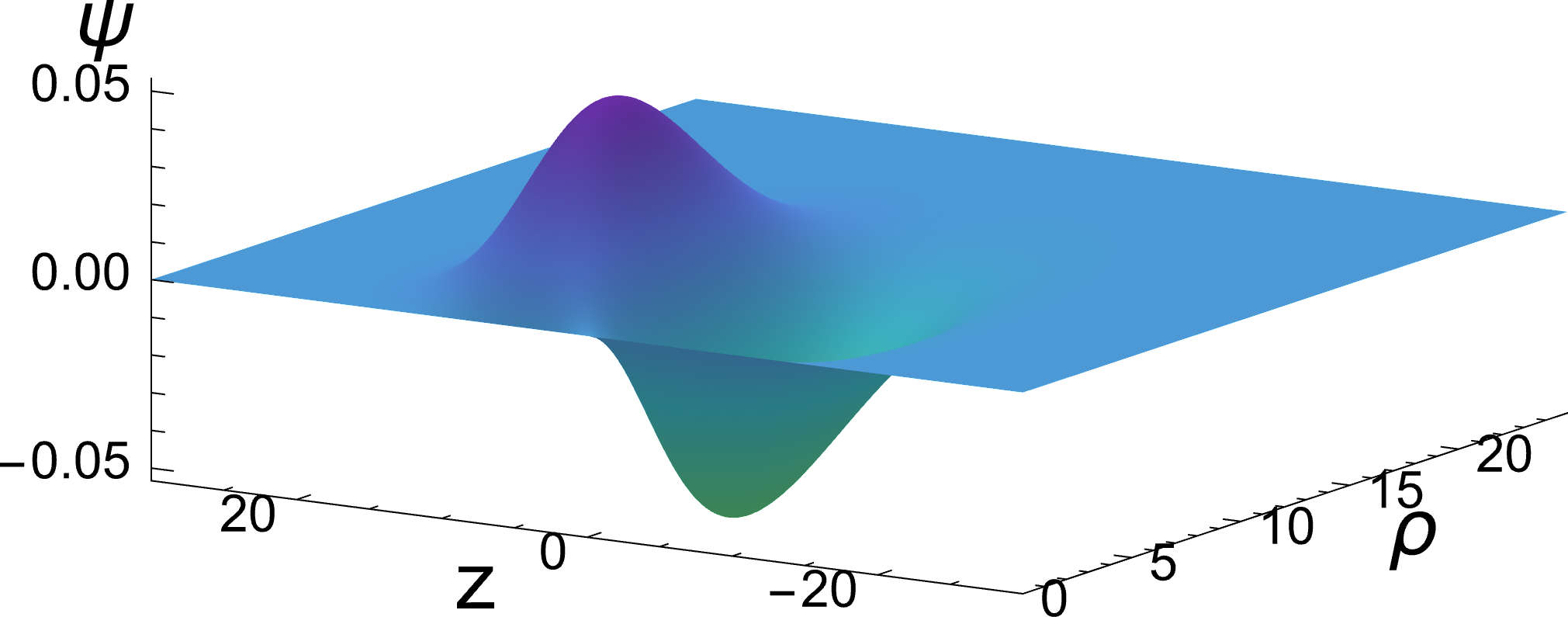}
    \hspace{0.01\textwidth}
  \includegraphics[width=0.31\textwidth]{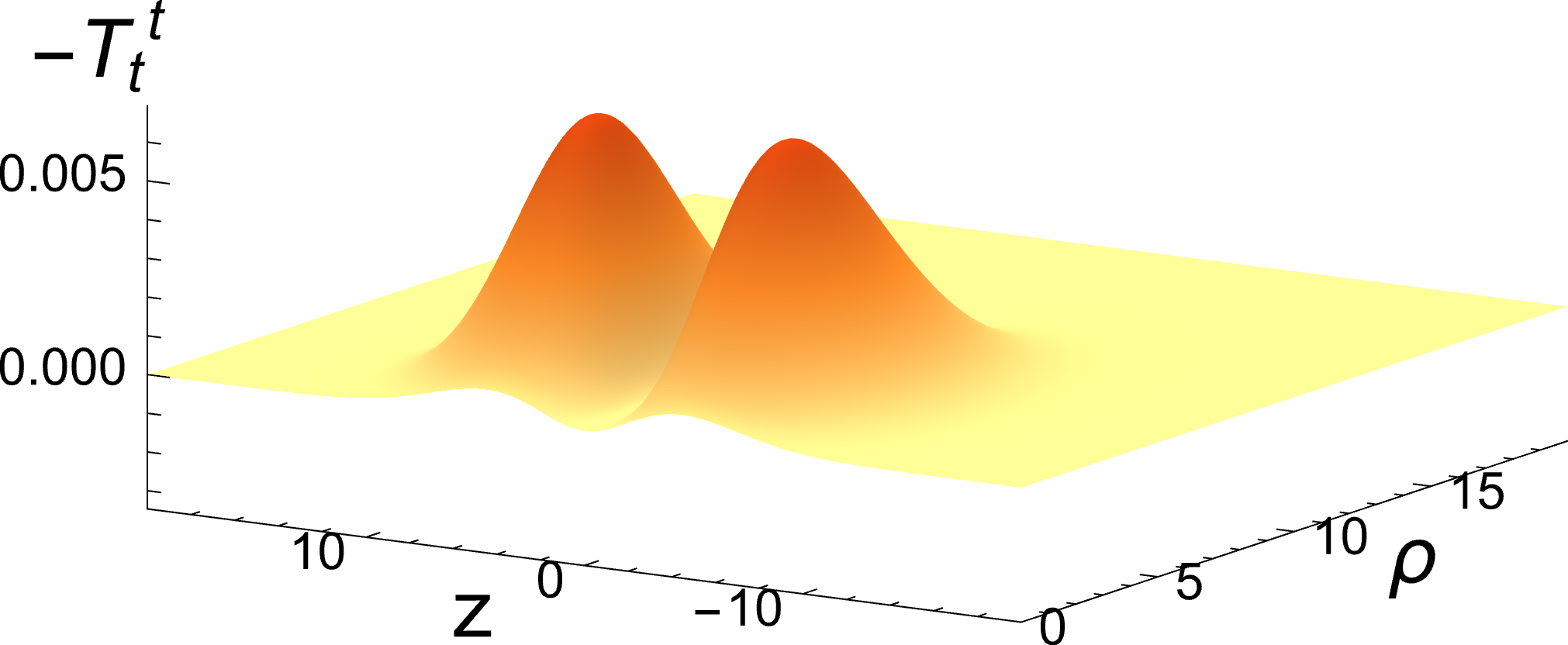}
    \hspace{0.01\textwidth}
  \includegraphics[width=0.31\textwidth]{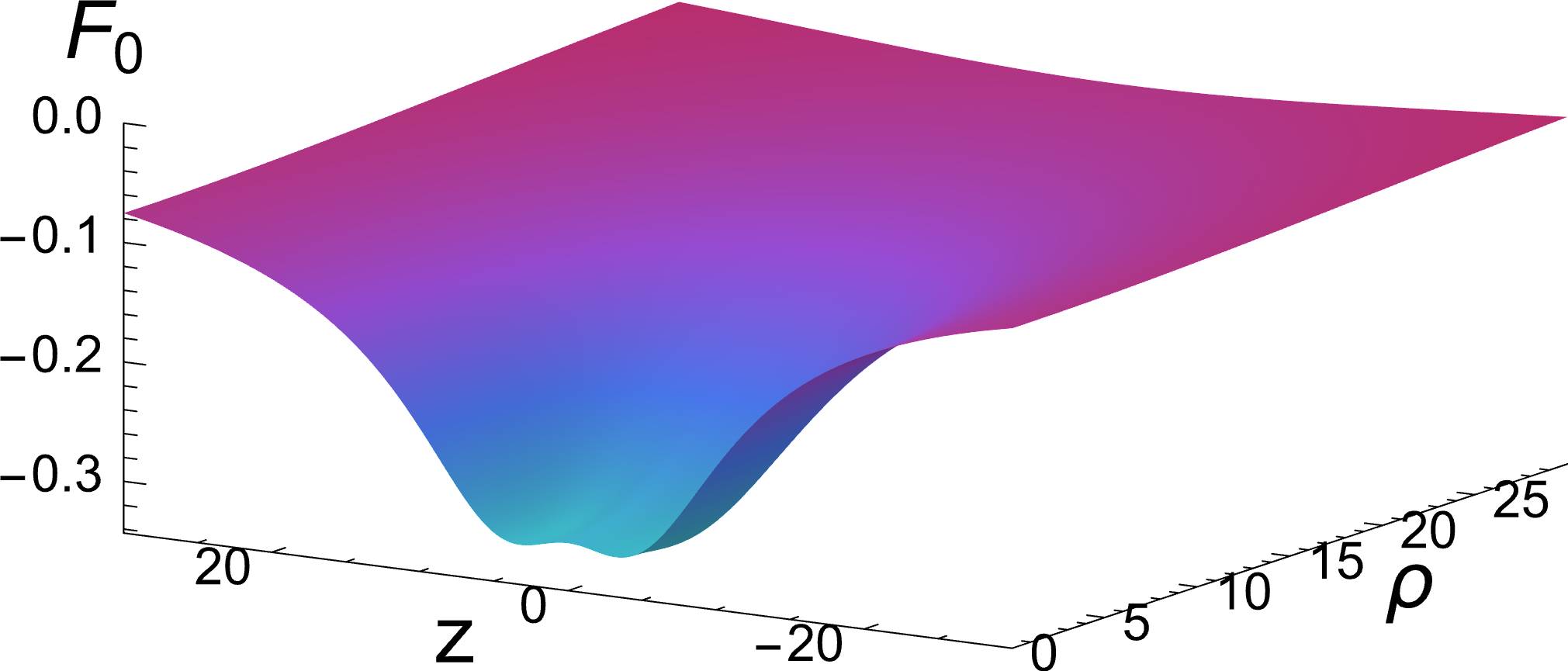}\\
  \vspace{0.05\textwidth}
  \includegraphics[width=0.31\textwidth]{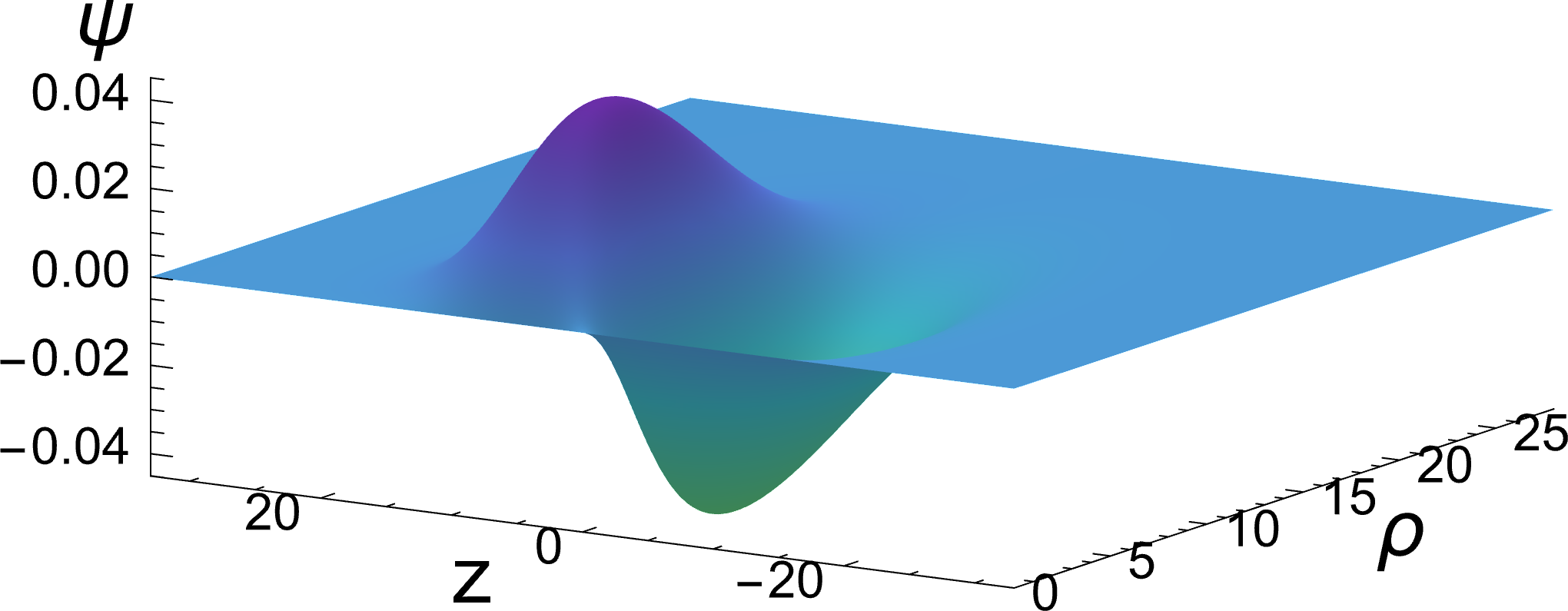}
    \hspace{0.01\textwidth}
  \includegraphics[width=0.31\textwidth]{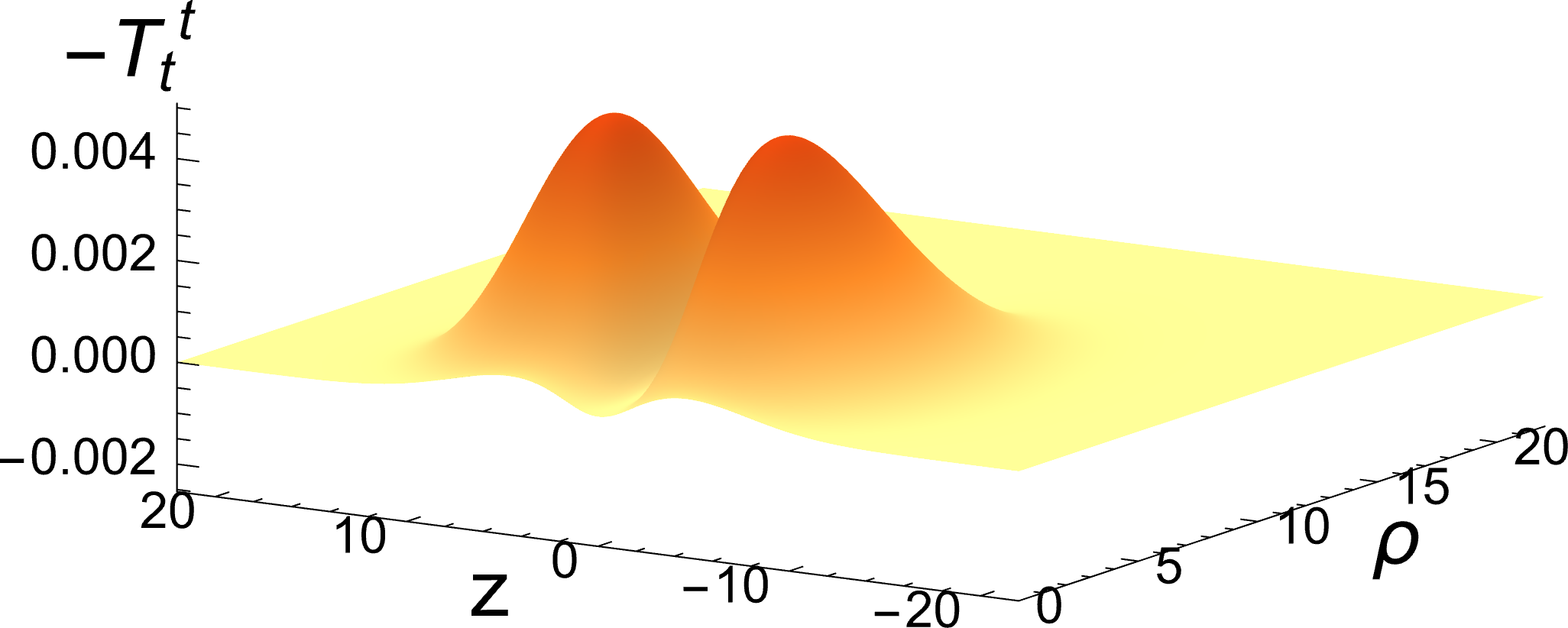}
    \hspace{0.01\textwidth}
  \includegraphics[width=0.31\textwidth]{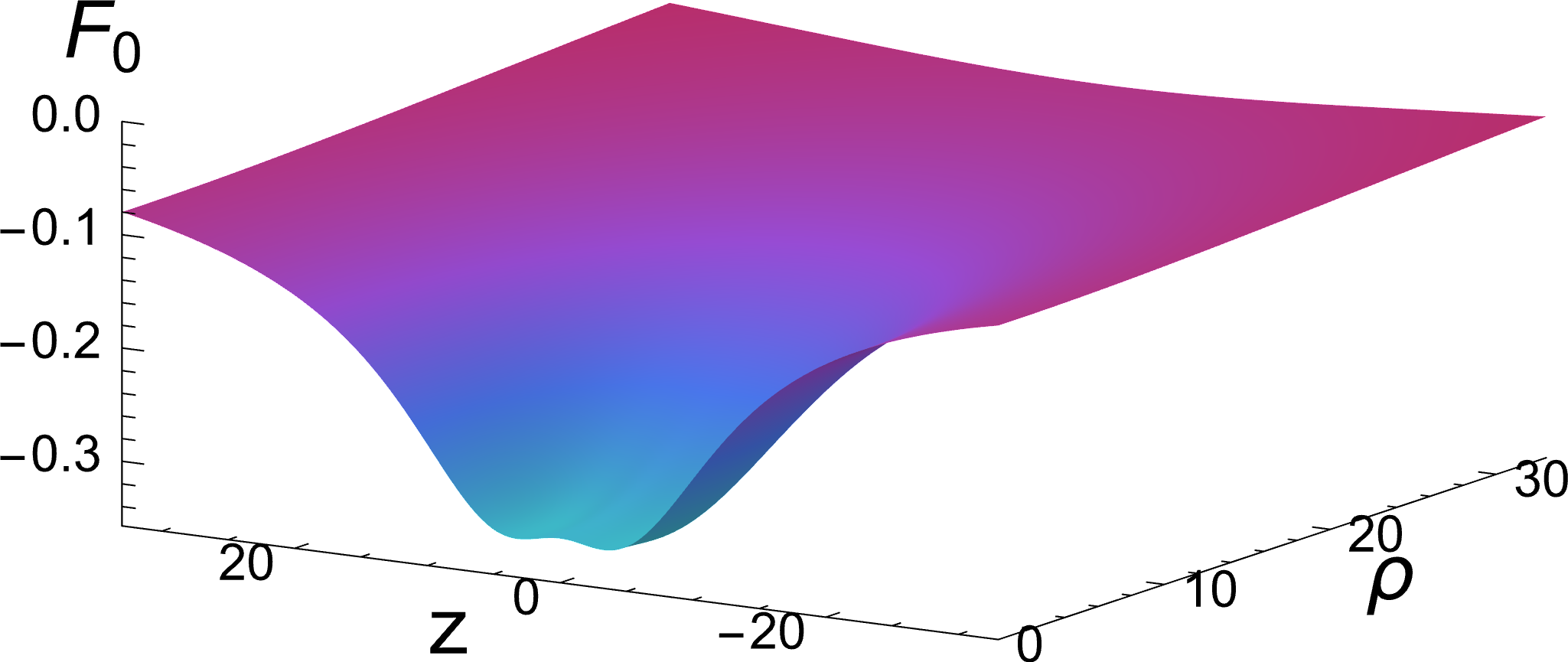}\\
  \vspace{0.05\textwidth}
  \includegraphics[width=0.31\textwidth]{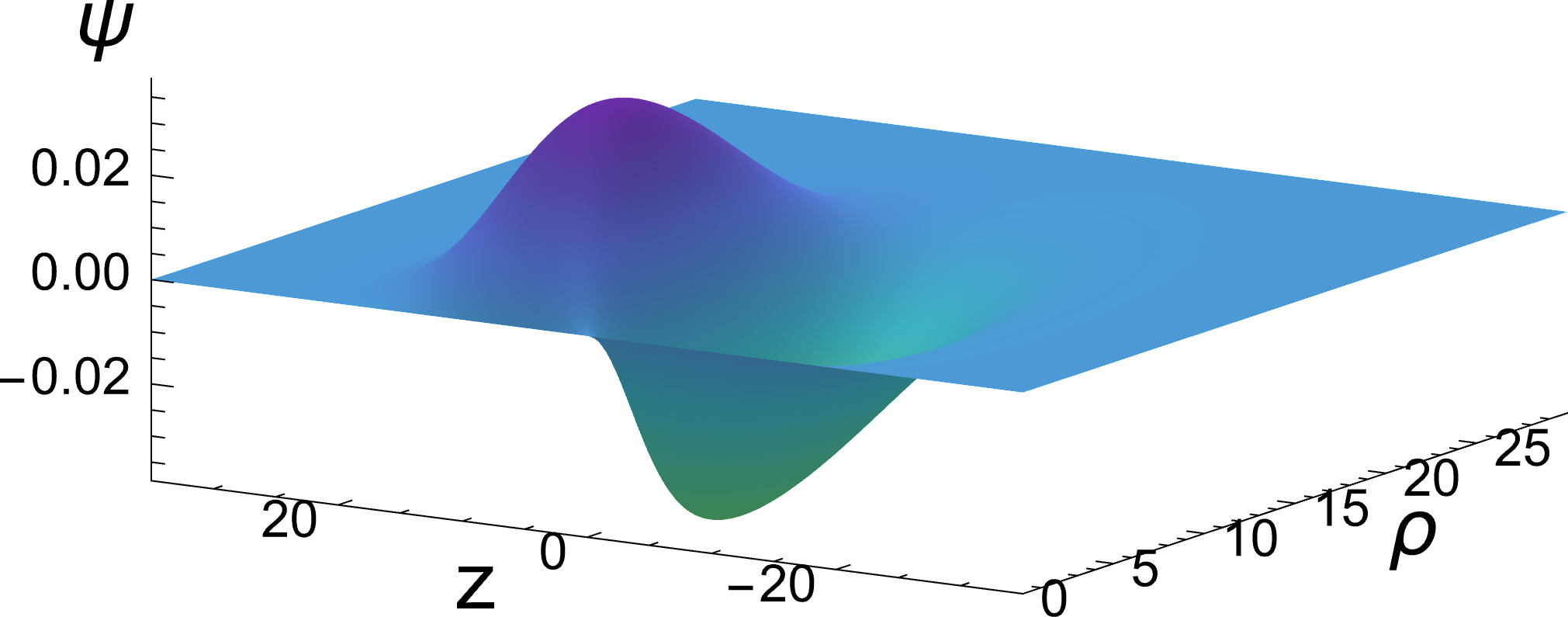}
    \hspace{0.01\textwidth}
  \includegraphics[width=0.31\textwidth]{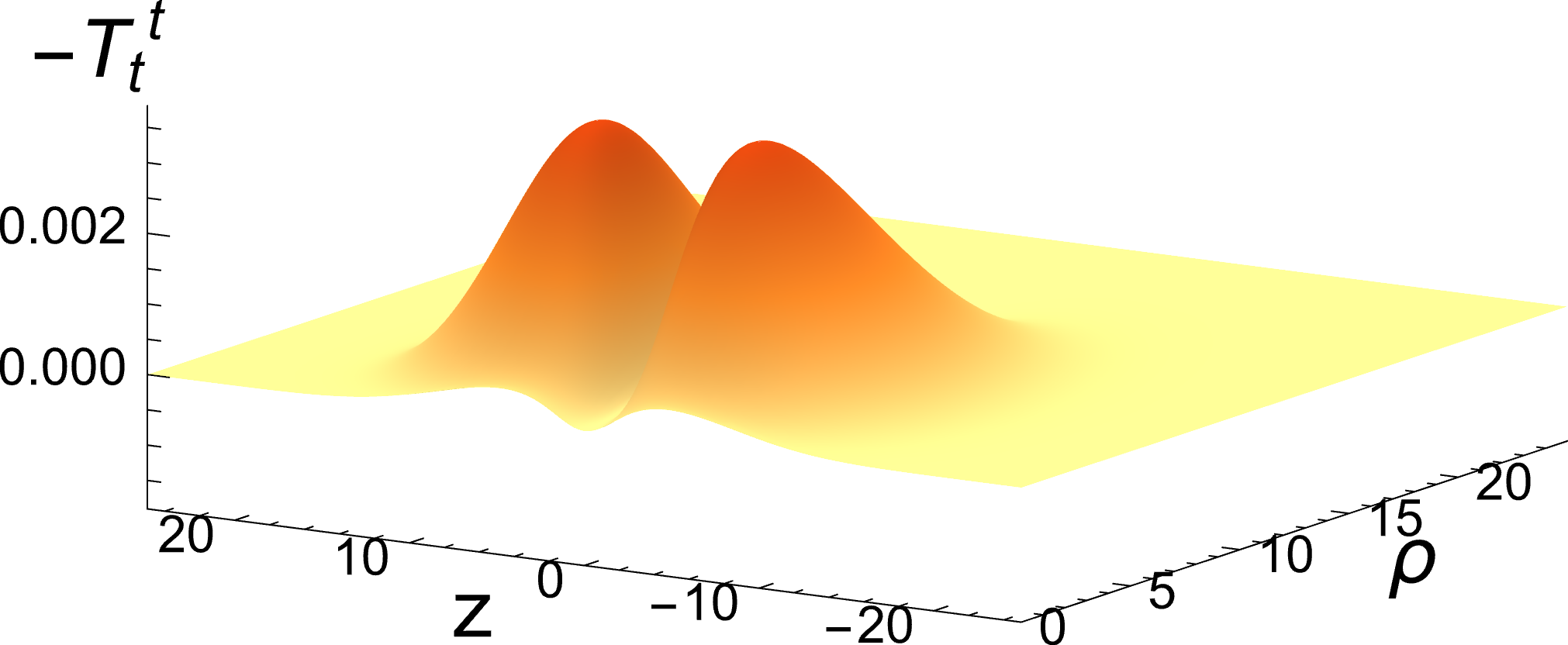}
    \hspace{0.01\textwidth}
  \includegraphics[width=0.31\textwidth]{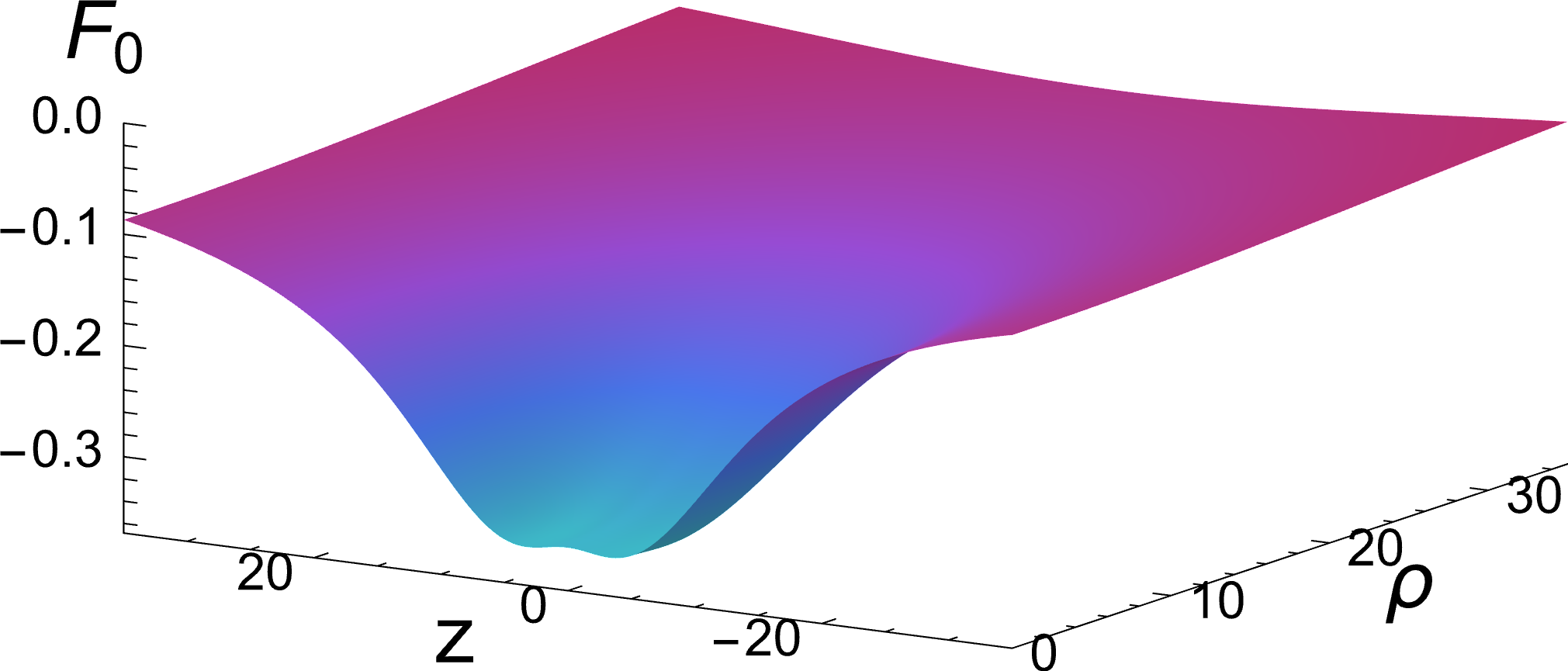}

  \caption{The scalar field amplitude $\psi$ (left column), energy density $-T_t^t$ (middle column), and metric function $F_0$ (right column) of 2sBSs with $\omega=0.85$ and $\lambda=0,30,60,100$ (from top to bottom).}
  \label{phiTttF0}
\end{figure}

We display the scalar field amplitude $\psi$, the energy density $-T_t^t$, and the metric function $F_0$ of 2sBSs in cylindrical coordinates in Fig.~\ref{phiTttF0}. The left column shows the amplitude $\psi$, which changes sign across the equatorial plane;
however, in contrast to the static case
\cite{Cunha:2022tvk},
 the presence of angular momentum causes $\psi$ to vanish on the $z$-axis.  Correspondingly, the energy density, shown in the middle column, exhibits two peaks symmetric about the equatorial plane; however, it does not vanish everywhere on the $z$-axis. As $\lambda$ increases, the repulsive self-interaction strengthens, and both the amplitude and energy density decrease. From the 3D plots of $\psi$ or $-T_t^t$, one can see that the matter distribution of 2sBSs takes the form of a double torus. This matter distribution provides a scalar environment capable of balancing the gravitational attraction between the two horizons of the 2sBHs, as we discuss in Sec.~\ref{sec:2sBHs}.

\section{Single hairy Kerr black holes}\label{sec:1sBHs}
\subsection{Ansatz and boundary conditions}
The 2sBHs we aim to construct in this work differ significantly from 1sBHs. To better highlight and understand the characteristics of 2sBHs in subsequent discussions, we first consider the case of 1sBHs. To begin with, we employ a metric ansatz similar to~\eqref{metric}:
\begin{eqnarray}
ds^2= -e^{2F_0(r,\theta)}N(r) dt^2 + 
e^{2F_1(r,\theta)}\left(\frac{dr^2}{N(r)}+r^2d\theta^2\right)+
e^{2F_2(r,\theta)}r^2\sin^2\theta\left(d\varphi-W(r,\theta)dt\right)^2
\,,
\end{eqnarray}
with $N(r)=1-{r_H}/{r}$. The event horizon is located at $r=r_H$; thus, 2sBS configurations correspond to $r_H=0$. The scalar field ansatz is the same as~\eqref{matter_ansatz}. Additionally, we introduce a new radial coordinate 
that is better suited
for numerically constructing 1sBHs:
\begin{eqnarray}
u=\sqrt{r^2-r_H^2}
\,.
\end{eqnarray}
Under this coordinate transformation, the event horizon is mapped to the origin $u=0$, and the metric becomes
\begin{align}
ds^2= & 
-\frac{e^{2F_0(u,\theta)} u^2H(u)}{v(u)}dt^2 + 
\frac{e^{2F_1(u,\theta)}}{H(u)}\left(du^2+H(u)v(u)\, d\theta^2\right)+
e^{2F_2(u,\theta)}v(u)\sin^2\theta\left(d\varphi-W(u,\theta)dt\right)^2
\,,
\end{align}
with
\begin{equation}
v(u) = u^2 + r_H^2\,,\qquad 
H(u) = \frac{\sqrt{v(u)}}{r_H+\sqrt{v(u)}}\,.
\end{equation}

Substituting the above ansatz into~\eqref{equek} yields a set of equations from which the matter and metric functions can be obtained numerically.

The equations of motion of 1sBHs are solved subject to the following boundary conditions:

\begin{itemize}
  \item At $u=0$, regularity requires
  \begin{align}
    \partial_u F_0|_{u=0}=\partial_u F_1|_{u=0}=\partial_u F_2|_{u=0}=0\,,\quad W|_{u=0}=\Omega_H  \,,\quad \psi|_{u=0}=0\,,
  \end{align}
  where $\Omega_H$ is the horizon angular velocity, which satisfies the synchronisation condition~\cite{Herdeiro:2014goa}
  \begin{align}
\Omega_H=\frac{\omega}{m}
\,.
\end{align}

  \item At infinity, asymptotic flatness requires that all functions vanish:
\begin{align}
F_0|_{u\to \infty}=F_1|_{u\to \infty}=F_2|_{u\to \infty}=W|_{u\to \infty}=\psi|_{u\to \infty}=0
\,.
\end{align}

  \item On the symmetry axis, regularity and axial symmetry require
  \begin{align}
\partial_\theta F_0|_{\theta=0,\pi}=\partial_\theta F_1|_{\theta=0,\pi}=\partial_\theta F_2|_{\theta=0,\pi}=\partial_\theta W|_{\theta=0,\pi}=0\,,\quad \psi|_{\theta=0,\pi}=0
\,.
\end{align}
Additionally, the constraint $F_1|_{\theta=0,\pi}=F_2|_{\theta=0,\pi}$ is imposed to avoid conical singularities.

  \item 
  In addition, the configurations possess 
  again a $\mathbb{Z}_2$ symmetry under which the scalar field amplitude 
  changes sign, the following conditions being imposed at $\theta=\pi/2$:
  \begin{align}
\partial_\theta F_0|_{\theta=\frac{\pi}{2}}=\partial_\theta F_1|_{\theta=\frac{\pi}{2}}=\partial_\theta F_2|_{\theta=\frac{\pi}{2}}=\partial_\theta W|_{\theta=\frac{\pi}{2}}=0
\,, \quad \psi|_{\theta=\frac{\pi}{2}}=0\,.
\end{align}
\end{itemize}

\subsection{Physical quantities}\label{quantities}
As in the horizonless case, the ADM mass and angular momentum can be read off directly from the asymptotic behavior~\eqref{asymptotic}. In addition, one can calculate the individual contributions of the horizon and the scalar hair to the ADM quantities via the Komar integrals: 
\begin{subequations}\label{mj}
  \begin{align}
M_H & =-\frac{1}{8\pi G}\oint_{\mathcal{H}}dS_{\mu\nu}\nabla^\mu\xi^\nu = 
\frac{1}{4G}\int_{0}^{\pi} d\theta \frac{1}{2}e^{F_0+F_2}\sqrt{v}\sin\theta\left( 1-2e^{2(-F_0+F_2)} v^2WW_{,uu}\sin^2\theta\right)\Big|_{u=0}
\,,
\\
J_H & =\frac{1}{16\pi G}\oint_{\mathcal{H}}dS_{\mu\nu}\nabla^\mu\eta^\nu = 
-\frac{1}{8G}\int_{0}^{\pi} d\theta 
\frac{1}{2} e^{-F_0+3F_2}v^{3/2}\sin^3\theta W_{,uu}\Big|_{u=0}
\,,\\\notag
\\\notag
M_\Psi & =-\frac{1}{4\pi G}\int_\Sigma dS_\mu(2T_\nu^\mu\xi^\nu-T\xi^\mu)\\
&= \frac{1}{G}\int_{0}^{\infty} du \int_{0}^{\pi} d\theta\frac{e^{-F_0+F_1+F_2}\sqrt{v}\sin\theta \psi^2}{u H}\left(2\omega v\left( \omega-mW \right) - \frac{e^{2F_0} u^2 H U}{\psi^2}\right)
\,,\\\notag
\\\notag
J_\Psi & =\frac{1}{8\pi G}\int_\Sigma dS_\mu\left(T_\nu^\mu\eta^\nu-\frac{1}{2}T\eta^\mu\right)\\
&= \frac{1}{2G}\int_{0}^{\infty} du \int_{0}^{\pi} d\theta \frac{e^{-F_0+2F_1+F_2}v^{3/2}m\sin\theta \psi^2}{uH}(\omega-mW)\,.
\end{align}
\end{subequations}
The total mass $M$ and angular momentum $J$ can thus be expressed as $M=M_H+M_\Psi$ and $J=J_H+J_\Psi$.
The Hawking temperature and event horizon area are given by
\begin{align}\label{thah}
T_H=\frac{\kappa}{2\pi}=\frac{1}{4\pi r_H}e^{F_0-F_1}\Big|_{u=0}
\,,\qquad
A_H=2\pi r_H^2 \int_{0}^{\pi} d\theta \sin\theta \left(e^{F_1+F_2}\right)\Big|_{u=0}
\,,
\end{align}
where $\kappa$ is the surface gravity, defined by $\kappa^2=-\frac{1}{2}(\nabla_a\chi_b)(\nabla^a\chi^b)|_{r_H}$, and $\chi=\xi + \Omega_H \eta$ is the null Killing vector on the horizon.

The above quantities for 1sBHs satisfy the Smarr relation
 \begin{align}
M=2T_H S + 2\Omega_H J_H + M_\Psi
\,,
\end{align}
where $S=A_H/(4G)$ is the entropy of the black hole. This relation is also used to verify the numerical accuracy of the solutions. Moreover, these physical quantities satisfy the first law of black hole thermodynamics:
 \begin{align}
dM=T_H dS + \Omega_H dJ
\,.
\end{align}

Furthermore, the presence of scalar hair can cause the horizon of 1sBHs to deform while maintaining spherical topology. The ratio of the following two quantities, the polar and equatorial circumferences, characterizes the degree of horizon deformation \cite{Delgado:2018khf}:
\begin{align}
 L_p= 2 r_H \int_{0}^{\pi} e^{F_1(r_H,\theta)} d\theta \,, \quad L_e=2\pi r_H e^{F_2(r_H,\pi/2)}\,.
\end{align}

\subsection{Numerical results}
The numerical method we employ to construct 1sBHs is essentially the same as that for 2sBSs, and we continue to adopt the rescalings~\eqref{rescalings}. However, while the accuracy of soliton solutions can typically be ensured by having a sufficient number of grid points near $x=1$, hairy BHs require sufficient grid points near both $x=0$ and $x=1$, i.e., in the vicinity of the horizon and spatial infinity, to guarantee accuracy. Therefore, we employ grids with more points in the radial direction, around $N_x \times N_\theta = 300 \times 50$, thereby ensuring that the relative error remains below $10^{-3}$.
\begin{figure}[h!]
  \centering
  \includegraphics[width=0.7\textwidth]{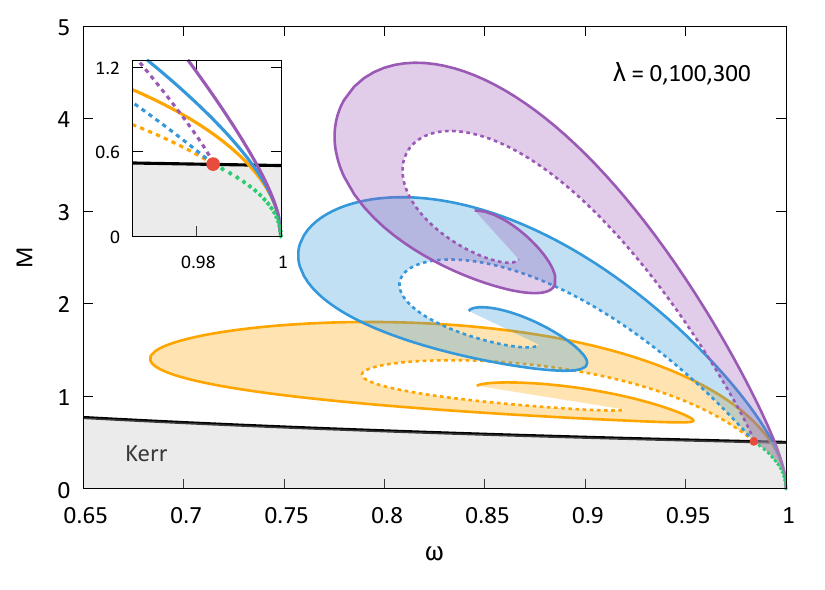}
  \caption{Domain of existence of 1sBHs with $\lambda=0,100,300$, represented by the shaded regions from bottom to top, respectively. The inset shows a magnified view near the Hod point, marked by a red dot.}
  \label{1wM}
\end{figure}

In Fig.~\ref{1wM}, we show the domain of existence of 1sBHs with $\lambda=0,100,300$. The black solid line indicates extremal Kerr BHs, which satisfy $M=1/(2\Omega_H)$. Extremal 1sBHs are represented by the dashed lines above the black solid line, exhibiting a spiral pattern similar to that of 2sBSs, which are represented by the solid lines. The green dotted line denotes the existence line for quadrupolar scalar clouds. Moreover, as clearly shown in the inset, the line of extremal 1sBHs and the existence line both intersect the extremal Kerr solution at a single point---the Hod point---marked by a red dot in the figure.

Each domain of existence of 1sBHs is bounded by the 2sBSs line, the existence line, and the line of extremal 1sBHs. As $\lambda$ increases, the maximum mass of both 2sBSs and extremal 1sBHs increases, while the range of the horizon angular velocity of 1sBHs decreases. However, for scalar clouds, the influence of higher-order terms in the potential $U(|\Psi|^2)$ is suppressed; therefore, the existence line does not vary with $\lambda$. The quadrupolar scalar hair considered here corresponds to the case 
$(l,m,n)=(2,1,0)$ in~\cite{Hod:2012px}, whose analytical estimate for the mass at the Hod point agrees well with our numerical 
results.

\begin{figure}[h!]
  \centering
  \includegraphics[width=0.49\textwidth]{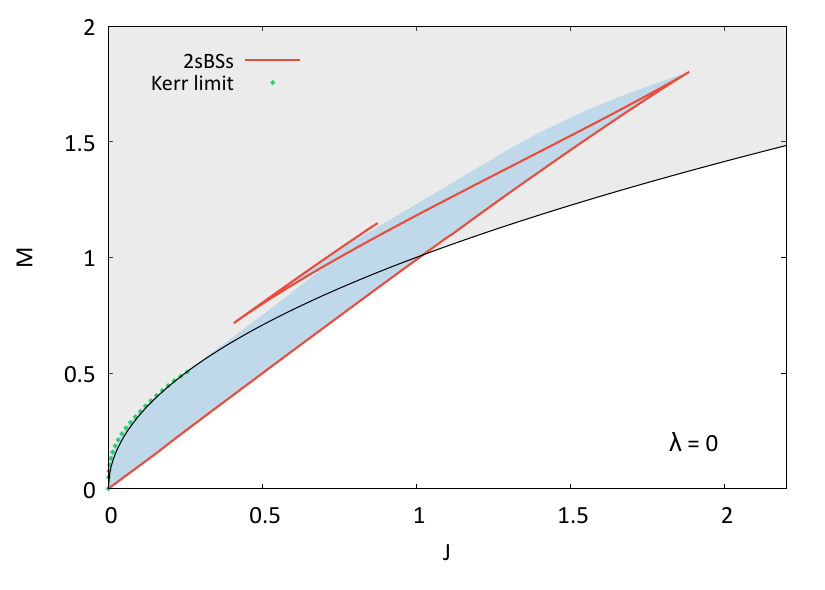}
  \includegraphics[width=0.49\textwidth]{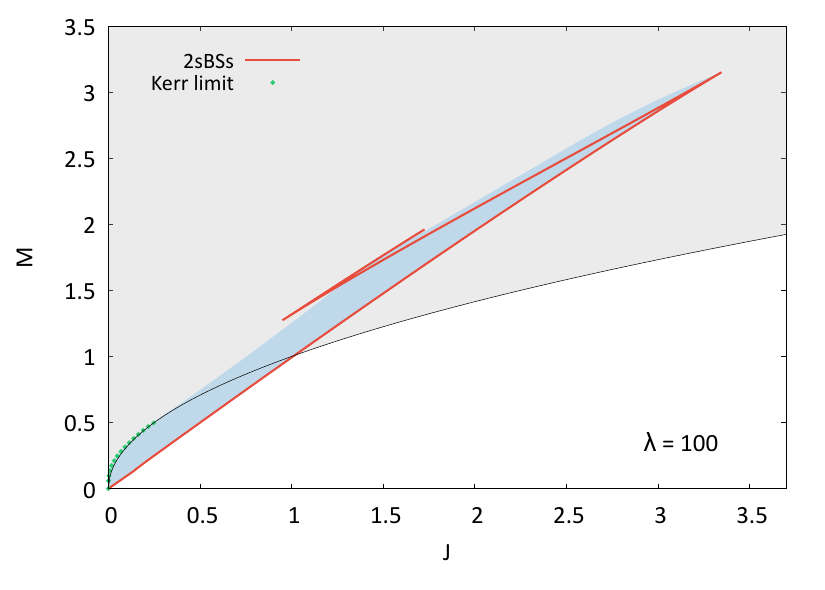}
  \includegraphics[width=0.49\textwidth]{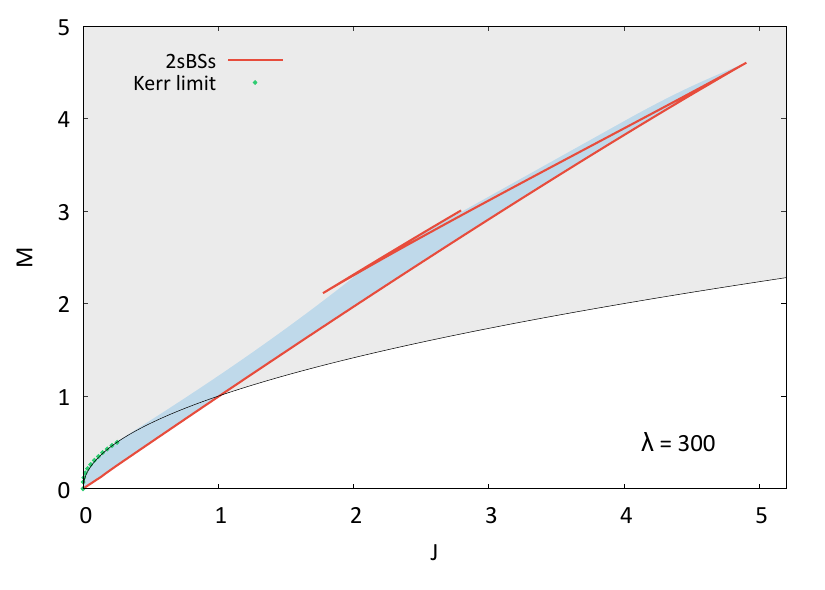}
  \includegraphics[width=0.49\textwidth]{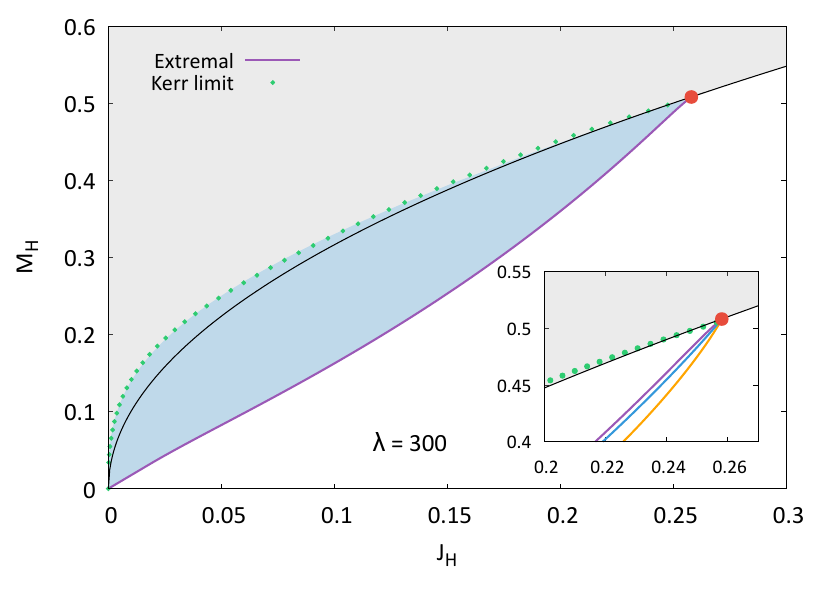}
  \caption{(Top and bottom left panels) ADM mass 
 $vs.$ ADM angular momentum for 1sBHs with $\lambda=0,100,300$. (Bottom right panel) Horizon mass $vs.$  horizon angular momentum for 1sBHs with $\lambda=300$. The inset shows a magnified view of these horizon quantities for extremal 1sBHs with $\lambda=0,100,300$.}
  \label{1JM}
\end{figure}

Fig.~\ref{1JM} displays the phase space 
of solutions
in terms of ADM and horizon quantities for 1sBHs with $\lambda=0, 100, 300$.
In each panel, the gray shaded region above 
the solid black line
represents Kerr BHs, bounded by extremal Kerr BHs indicated by the black solid line. The red solid line represents 2sBSs with the corresponding $\lambda$, while the green dotted line denotes the Kerr limit, $i.e.$, the existence line.
The blue shaded region indicates the domain of existence for 1sBHs, obtained by extrapolating from discrete sets of thousands of numerical solutions.
Both the upper and lower bounds of $M$ and $J$ for 1sBHs are determined by 2sBSs. Moreover, since 1sBHs can approach 2sBSs arbitrarily closely by reducing the horizon radius, there is always a subset of solutions within their existence domain that violate the Kerr bound. Furthermore, as $\lambda$ increases, the existence domain of 1sBHs in phase space extends along the direction of increasing $M$ and $J$ of 2sBSs. However, the domain of 1sBHs violating the Kerr bound remains nearly unchanged, with its upper boundary always located around $(J, M) \sim (1, 1)$.

The bottom right panel shows the phase space in terms of the horizon quantities $M_H$ and $J_H$ for 1sBHs with $\lambda=300$. The purple line represents extremal 1sBHs, which, together with the existence line, bound the region of 1sBHs. The origin in this diagram corresponds to 2sBSs. As $J_H$ increases, $M_H$ of both extremal 1sBHs and the existence line increases, eventually reaching the Hod point, where they intersect with extremal Kerr, marked by the red point in the panel. Similar to the results obtained in~\cite{Herdeiro:2015tia}, the values of $M_H$ and $J_H$ for 1sBHs are always smaller than those at the Hod point. Varying $\lambda$ only slightly changes $M_H$ and $J_H$ of extremal 1sBHs near the Hod point, as shown in the inset, where the orange and blue lines represent extremal 1sBHs with $\lambda=0$ and $100$, respectively. Therefore, the values of $M_H$ and $J_H$ for 1sBHs always range from zero to the Hod point, independent of $\lambda$.

This demonstrates that although the introduction of self-interactions can significantly increase the ADM mass $M$ and angular momentum $J$ of 1sBHs, the horizon quantities $M_H$ and $J_H$ are always bounded above by the Hod point. In other words, for quadrupolar scalar hair, we obtain the same result as for the fundamental scalar hair studied in~\cite{Herdeiro:2015tia}: self-interactions enable 1sBHs to be \textit{hairier}---i.e., to possess scalar hair with larger mass---but not \textit{heavier}---i.e., the horizon mass cannot exceed that at the Hod point.

\begin{figure}[h!]
  \centering
  \includegraphics[width=0.49\textwidth]{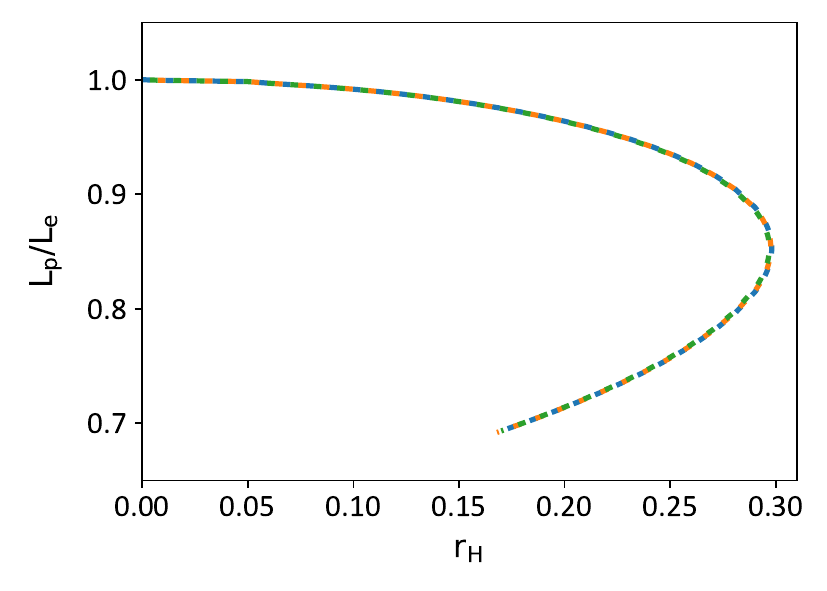}
  \includegraphics[width=0.49\textwidth]{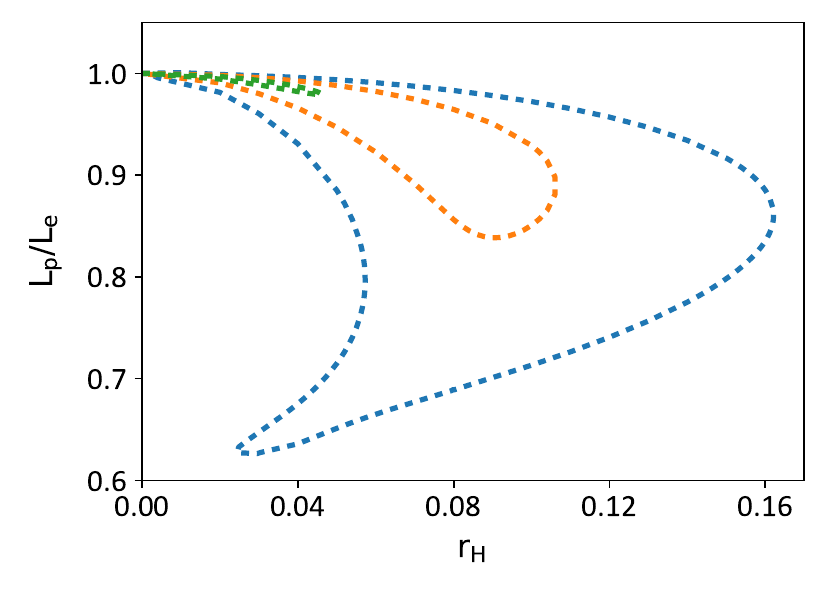}
  \caption{The horizon deformation $L_p/L_e$ of 1sBHs as a function of $r_H$ for $\lambda=0$, $100$, and $300$ shown as blue, orange, and green dashed lines. The left and right panels correspond to $\omega=0.99$ and $\omega=0.78$, respectively.}
  \label{lple}
\end{figure}

Fig.~\ref{lple} shows the horizon deformation of 1sBHs with different values of $\lambda$ and $\omega$. For the case $\omega=0.99$, which is close to the bound state limit, the scalar field is diluted and the higher-order terms in the self-interaction are suppressed; consequently, $L_p/L_e$ hardly changes as $\lambda$ increases. In contrast, for the case $\omega=0.78$, as shown in the right panel, the 1sBHs deviate significantly from the vacuum Kerr solution and the system is dominated by the scalar field. Moreover, as $\lambda$ increases, the maximum size of the 1sBHs gradually decreases, so that each 1sBH can be regarded to some extent as a test particle immersed in the scalar background; in this limit the horizon experiences negligible deformation, and $L_p/L_e$ approaches unity.

\subsection{An effective model $vs.$ numerical results}

\begin{figure}[h!]
  \centering
  \includegraphics[width=0.32\textwidth]{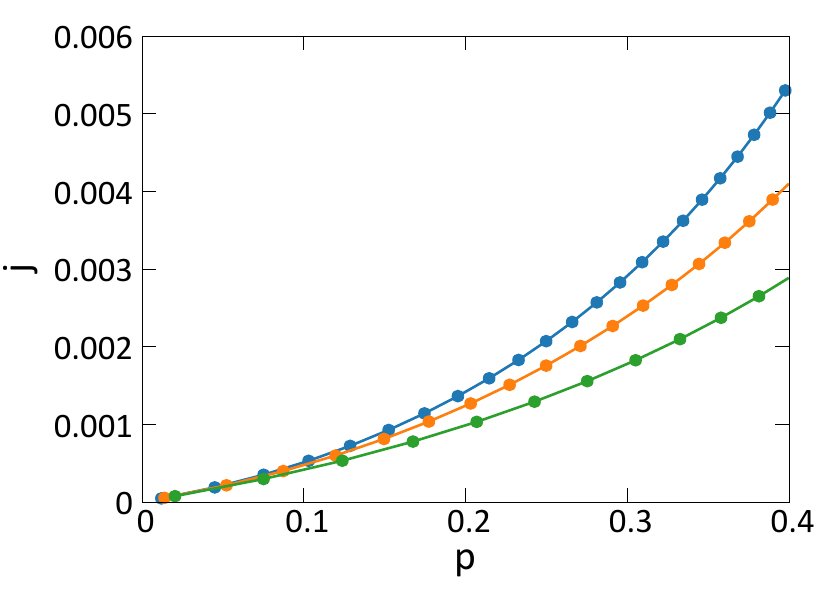}
  \includegraphics[width=0.32\textwidth]{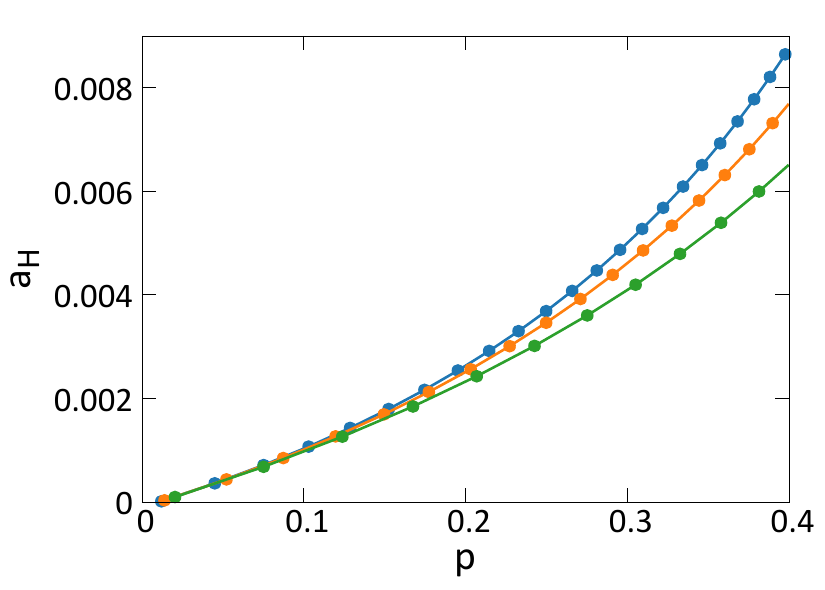}
  \includegraphics[width=0.32\textwidth]{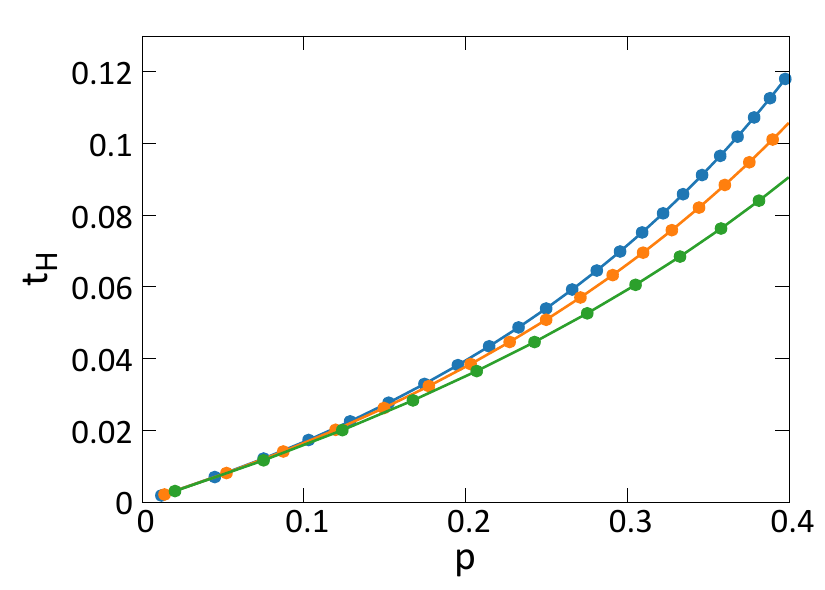}
  \includegraphics[width=0.32\textwidth]{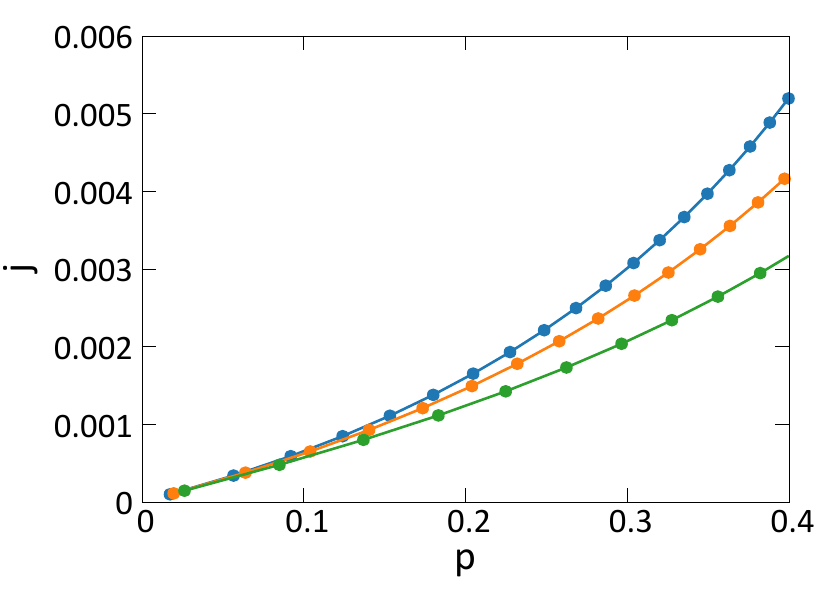}
  \includegraphics[width=0.32\textwidth]{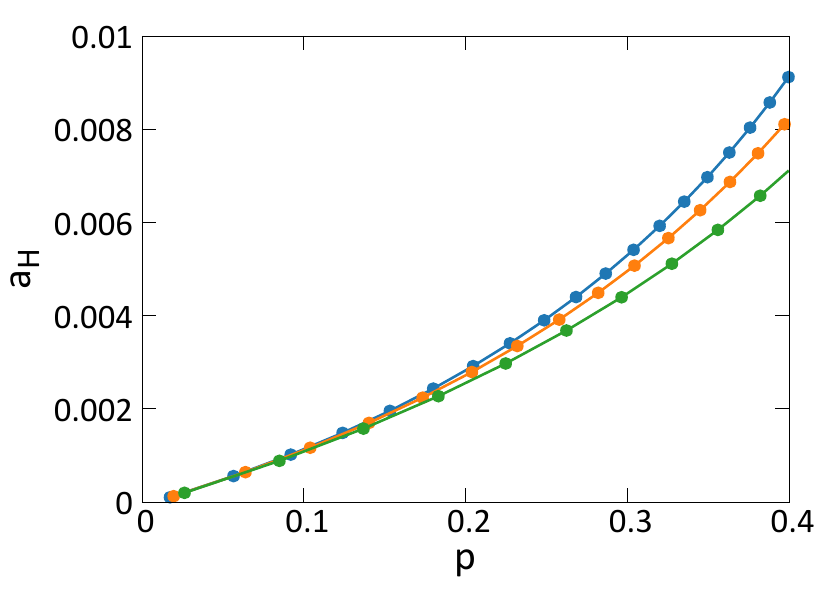}
  \includegraphics[width=0.32\textwidth]{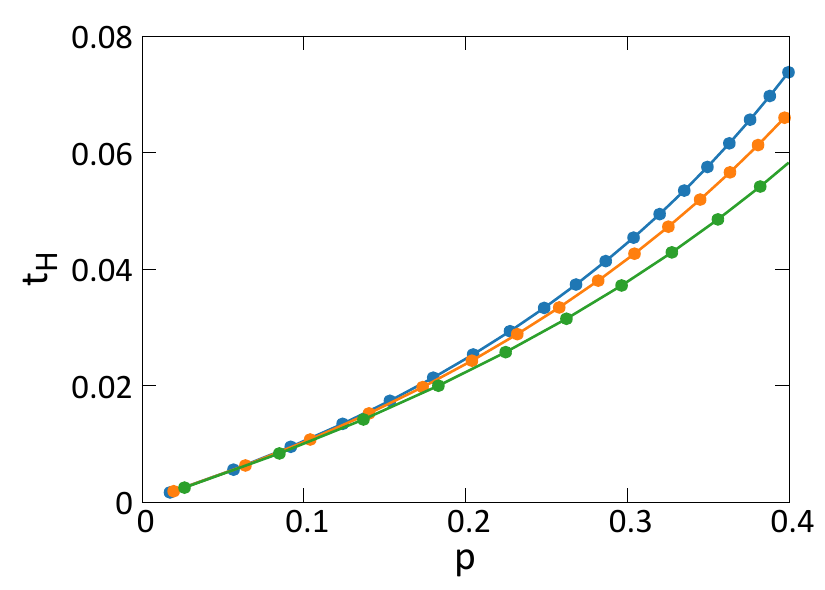}
  \caption{Relative errors of $j$, $a_H$, and $t_H$ as functions of $p$. The top and bottom rows correspond to 1sBHs with $r_H=0.2$ and $r_H=0.24$, while the blue, orange, and green lines correspond to $\lambda=0$, $100$, and $300$. Only the dots correspond to actual numerical solutions, with the connecting lines serving as guides to the eye.}
  \label{effective}
\end{figure}

\begin{figure}[h!]
  \centering
  \includegraphics[width=0.32\textwidth]{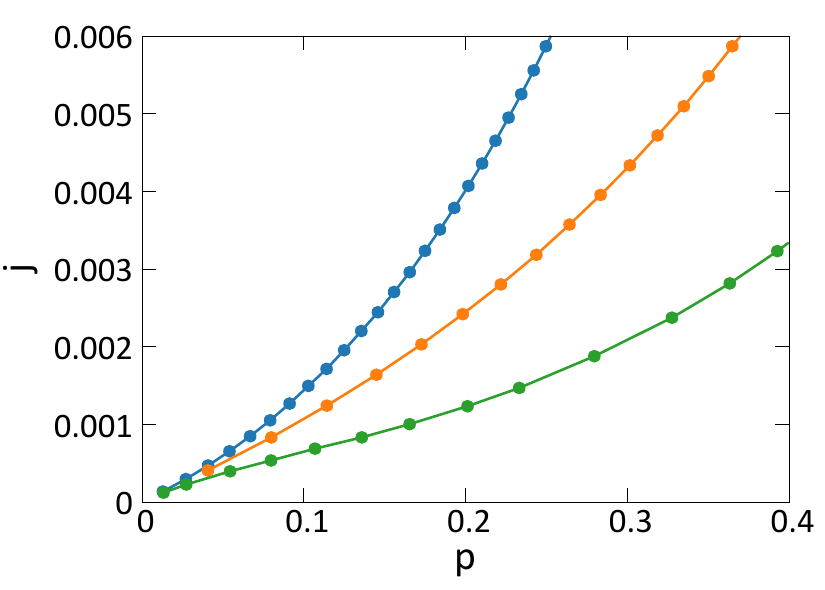}
  \includegraphics[width=0.32\textwidth]{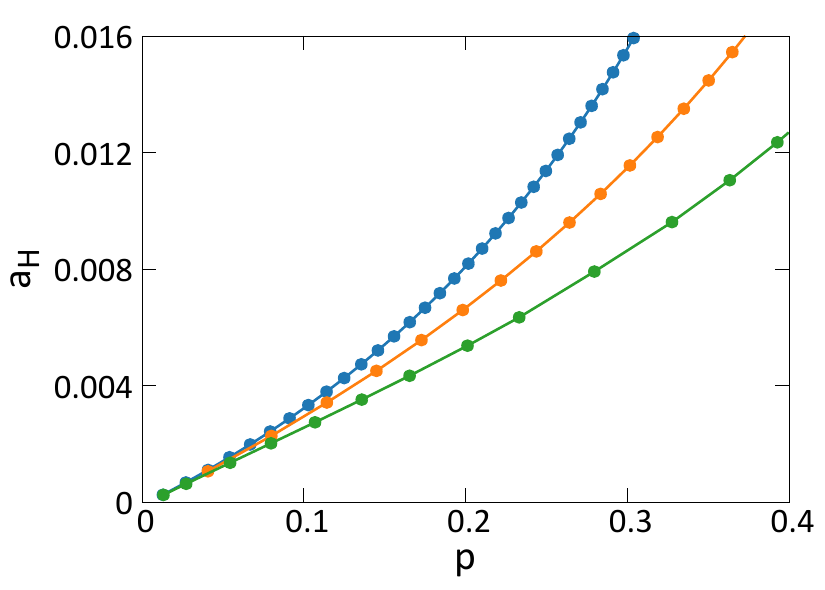}
  \includegraphics[width=0.32\textwidth]{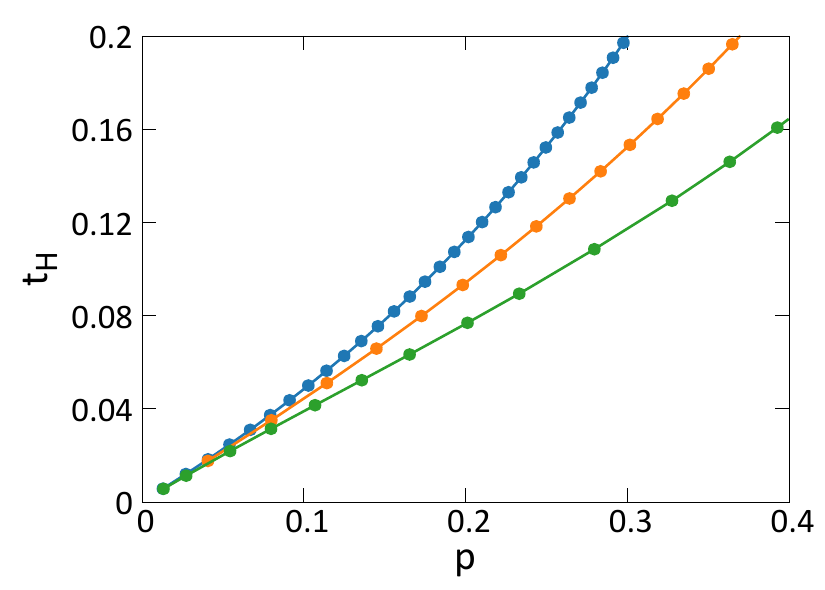}
  \includegraphics[width=0.32\textwidth]{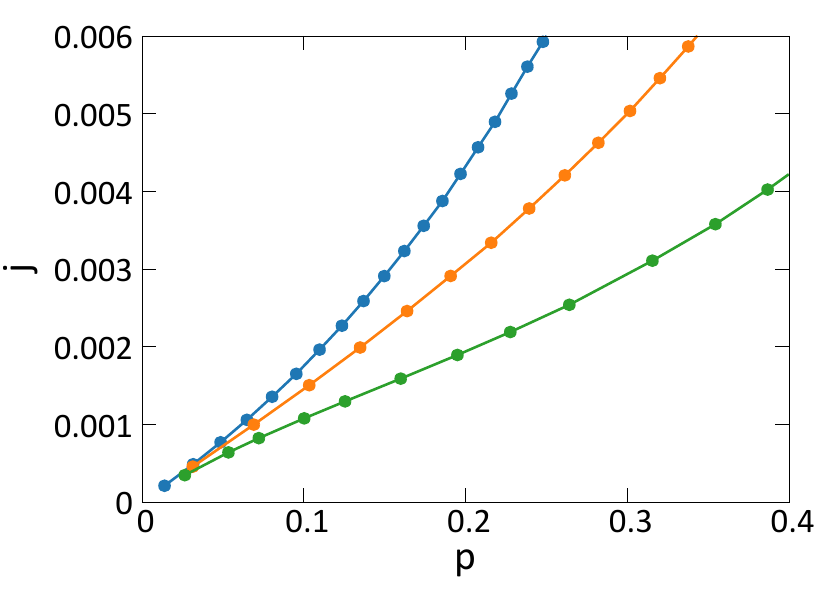}
  \includegraphics[width=0.32\textwidth]{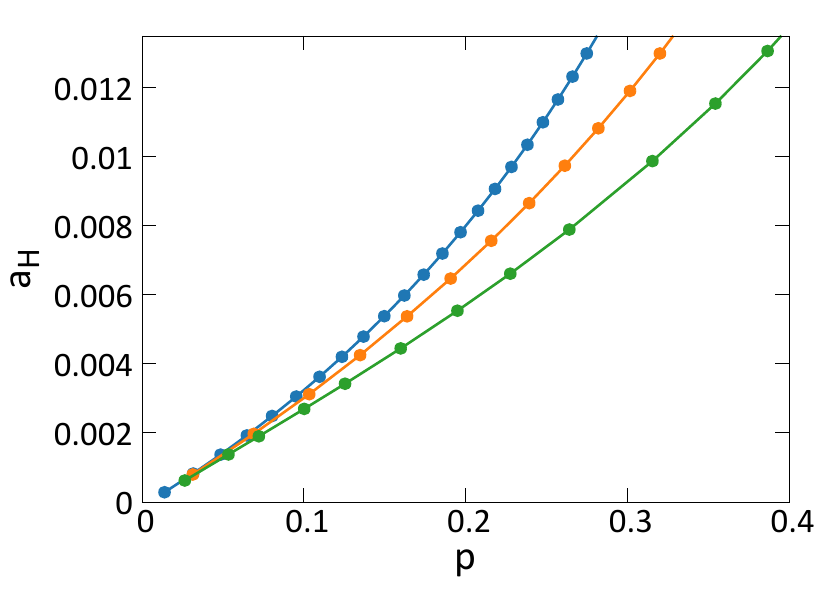}
  \includegraphics[width=0.32\textwidth]{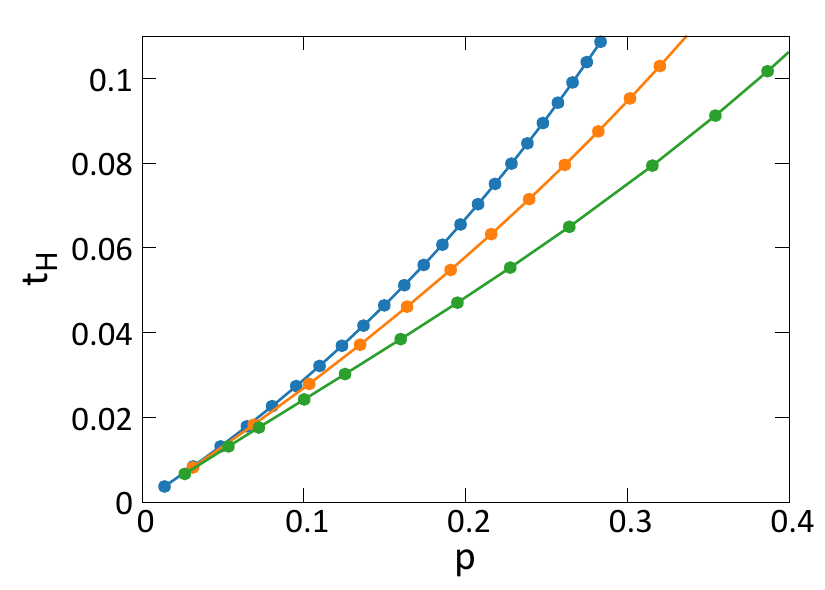}
  \caption{Same as Fig.~\ref{effective} for single Kerr BHs with fundamental self-interacting scalar hair.}
  \label{effective_even}
\end{figure}

Hairy BHs are highly nonlinear systems, and their exact solutions can 
only
be obtained through numerical methods. However, Ref.~\cite{Herdeiro:2017phl} proposed an analytical effective model to describe Kerr BHs with scalar hair
which captures some of their basic properties. This model yields results that are highly consistent with numerical solutions in the weak hair regime. This effective model was further generalized in ~\cite{Brihaye:2018woc}
for higher dimensions and a different spacetime asymptotics. Following the approach in~\cite{Herdeiro:2017phl}, we can define two quantities related to mass and angular momentum, respectively, which characterize the ``hairiness'' of 1sBHs:
\begin{align}
  p=\frac{M_\Psi}{M}\,,\qquad q=\frac{J_\Psi}{J}\,,
\end{align}
where the limits $p=0=q$ and $p=1=q$ correspond to vacuum Kerr and 2sBSs, respectively. Furthermore, we introduce the following physical quantities normalized by the ADM mass:
\begin{align}
  j=\frac{J}{M^2}\,,\quad a_H=\frac{A_H}{16\pi M^2}\,, \quad \omega_H=\Omega_H M\,,\quad t_H= 8\pi M T_H\,.
\end{align}
Then, taking $p$ and $\omega_H$ as independent parameters, the remaining physical quantities can be expressed as
\begin{align}
  q&=p\frac{1+4(1-p)^2\omega_H^2}{p+4(1-p)^2\omega_H^2}\,,\qquad j=\frac{p+4(1-p)^2\omega_H^2}{\omega_H(1+4(1-p)^2\omega_H^2)}\,,
  \\
 a_H&=\frac{(1-p)^2}{1+4(1-p)^2\omega_H^2}\,,\qquad t_H=\frac{1-4(1-p)^2\omega_H^2}{1-p}\,.
\end{align}

Fig.~\ref{effective} shows the relative errors of the physical quantities obtained from the above approximations, defined as $|1 - \text{Effective model results}/\text{Numerical results}|$, as functions of $p$ for given $r_H$,
the varying parameter being $\Omega_H$. 
When $p < 0.1$, the effective model yields highly accurate results; in particular, the relative errors of $j$ and $a_H$ are almost all below $10^{-3}$, while the error of $t_H$ is larger but still achieves satisfactory accuracy when $r_H$ is larger, i.e., $r_H = 0.24$, with relative errors below $10^{-2}$. Furthermore, when $p < 0.1$, self-interactions have almost no effect on the errors. As $p$ increases, the scalar field contributes a larger fraction of the total energy of the system, and the errors of the effective model grow accordingly. Interestingly, however, the introduction of self-interactions can significantly improve the applicability of the effective model in the regime of larger $p$: increasing the coupling $\lambda$ reduces the relative errors of the various physical quantities to some extent. 

The error reduction induced by self-interactions raises a natural question. Is this behavior tied to the quadrupolar nature of the scalar
hair in the 1sBHs? To address this, we have examined the effective model for single Kerr BHs with fundamental self-interacting scalar hair. The corresponding relative errors are shown in Fig.~\ref{effective_even}. Self-interactions likewise reduce the errors of the effective model in this case, and the reduction is in fact more pronounced than for the 1sBHs with quadrupolar scalar hair. Nevertheless, the overall errors of the effective model remain somewhat smaller in the 1sBHs with quadrupolar scalar hair than in the single Kerr BHs with fundamental scalar hair.

A natural explanation for the error reduction induced by self-interactions can be inferred from Figs.~\ref{effective} and \ref{effective_even}. For small $p$, e.g. $p < 0.1$, the solutions are close to the vacuum Kerr limit and the contribution of higher-order scalar terms in the potential (\ref{potential}) is negligible. The errors of the effective model are therefore essentially insensitive to the coupling $\lambda$. For larger $p$, $e.g.$ $p > 0.2$, the higher-order scalar terms become non-negligible, but the repulsive self-interactions also dilute the scalar matter distribution. At fixed $p$, increasing $\lambda$ leads to a more diluted scalar configuration, thereby suppressing its impact on the accuracy of the effective model.

\section{Double hairy Kerr black holes}\label{sec:2sBHs}
\subsection{The Bach-Weyl metric}
\begin{figure}[h!]
  \centering
  \includegraphics[width=0.15\textwidth]{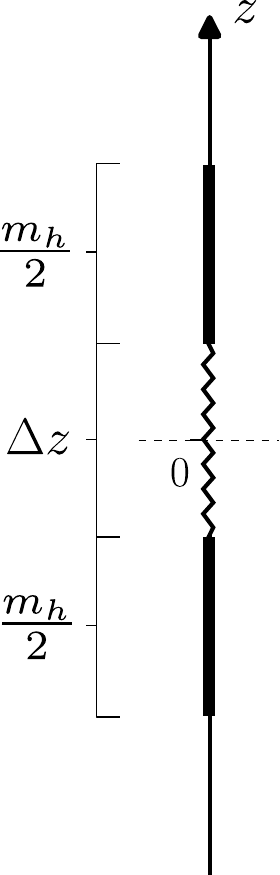}
\hspace{3cm}
  \includegraphics[width=0.15\textwidth]{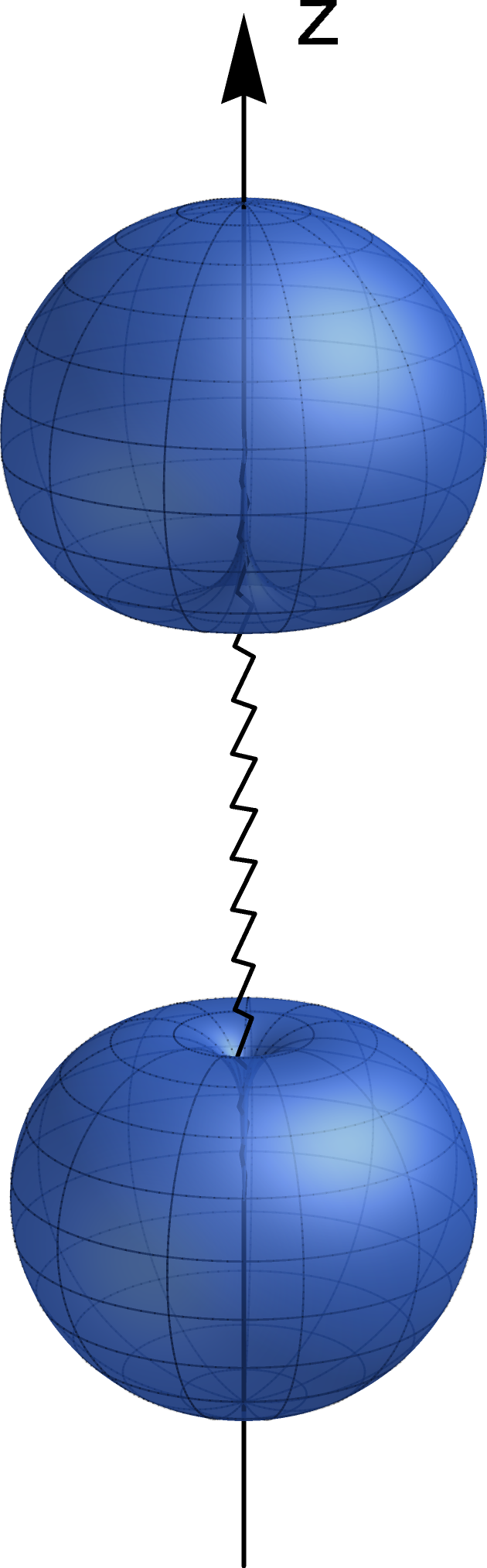}
  \caption{The rod structure of Bach-Weyl metric (left panel). The $z$-axis is composed by several rods: (i) two semi-infinite space-like rods, $[-\infty,-m_h/2-\Delta z/2]$ and $[m_h/2+\Delta z/2,\infty]$, extending to asymptotic infinity; (ii) two time-like rods (event horizons), $[-m_h/2-\Delta z/2,-\Delta z/2]$ and $[\Delta z/2,m_h/2+\Delta z/2]$; (iii) a space-like rod, $[-\Delta z/2,\Delta z/2]$, with a conical singularity. Visualization of the geometry (right panel) shows topologically spherical horizons balanced by a strut, represented by the zigzag line segment between them.
}
  \label{BW}
\end{figure}

Before delving into the 2sBHs, we would like to give a brief 
review of
the Bach-Weyl (BW) metric, which describes two collinear Schwarzschild BHs along the $z$-axis. First, consider a static axisymmetric metric,
\begin{eqnarray}\label{BW_metric}
ds^2= -e^{2V(\rho,z)}dt^2 + 
e^{-2V(\rho,z)}\left[
e^{2K(\rho,z)}(d\rho^2+dz^2)+
\rho^2d\varphi^2
\right]
\,.
\end{eqnarray}
Substituting the metric into vacuum Einstein equations and combining the equations appropriately yields the following equations for the metric functions $V$ and $K$,
\begin{subequations}\label{equvk}
\begin{align}
\label{equv}
&
\left(
\frac{\partial^2}{\partial\rho^2}+
\frac{1}{\rho}\frac{\partial}{\partial\rho}+
\frac{\partial^2}{\partial z^2}
\right)V=0\,, \\
\label{equk1}
&
K_{,\rho}=\rho(V_{,\rho}^2-V_{,z}^2)\,, \\
\label{equk2}
&
K_{,z}=2\rho V_{,\rho}V_{,z}\,.
\end{align}
\end{subequations}
Clearly, the equation (\ref{equv}) indicates that the metric function $V$ is identical to a Newtonian potential sourced by axisymmetric matter distribution, which is known as the \textit{rod structure}. In 1922, Bach and Weyl proposed the solution corresponding to two rods with a separation of $\Delta z$~\cite{Bach1922}:
\begin{subequations}\label{weyl}
\begin{align}
\label{equ3.3a}
&
e^{2V}=\frac{(r_1+r_2-m_h)(r_3+r_4-m_h)}{(r_1+r_2+m_h)(r_3+r_4+m_h)}\,, \\
\notag
&
e^{2K}=\left( \frac{\Delta z}{\Delta z + m_h} \right)^2
\left( \frac{(r_1+r_2)^2-m_h^2}{4r_1r_2} \right)
\left( \frac{(r_3+r_4)^2-m_h^2}{4r_3r_4} \right)
\\
\label{equ3.3b}
&\qquad \quad
\left( \frac{(\Delta z + m_h)r_1+ (\Delta z + 2m_h)r_2 - m_hr_4}{\Delta z~r_1+ (\Delta z + m_h)r_2 - m_hr_3} \right)^2
\,, 
\end{align}
\end{subequations}
where
\begin{equation}\label{r1234}
\begin{aligned}
&
r_1=\sqrt{\rho^2+\left( z - \frac{\Delta z}{2} - m_h \right)^2}
\,,\quad 
r_2=\sqrt{\rho^2+\left( z - \frac{\Delta z}{2}\right)^2}
\,, 
\\
&
r_3=\sqrt{\rho^2+\left( z + \frac{\Delta z}{2} \right)^2}
\,,\quad 
r_4=\sqrt{\rho^2+\left( z + \frac{\Delta z}{2} + m_h\right)^2}
\,.
\end{aligned}
\end{equation}
The ADM mass of the space-time is $m_h$, which is also twice the length of each rod. Therefore, the $z$-axis is divided into several intervals, as shown in Fig.~\ref{BW}. Due to their mutual gravitational attraction, the two BHs require an additional force to counterbalance gravity. A conical singularity situated between the BHs serves as a repulsive force that offsets the attraction, resulting in a deviation from the usual $2\pi$ periodicity. To characterize the properties of the conical singularity, we introduce the following quantity:
\begin{eqnarray}
\delta=2\pi\left( 1-\lim_{\rho \rightarrow 0}\sqrt{\frac{g_{\varphi\varphi}}{\rho^2g_{\rho\rho}}} \right)
\,.
\end{eqnarray}
Thus, for the Bach-Weyl solution (\ref{weyl}), $\delta=-2\pi m_h^2/(\Delta z^2+2m_h\Delta z)<0$. 
A conical excess occurs when $\delta<0$, acting like a strut that pushes the two BHs apart and counterbalances their gravitational attraction. Conversely, a conical deficit arises when $\delta>0$, which can be visualized as a segment along the $z$-axis behaving like a stretched string. In principle, the conical singularity between the BHs can be eliminated by rescaling the angular coordinate $\varphi$ such that $\delta=0$. However, this procedure transfers the singularity to two semi-infinite space-like rods extending from the event horizons outward along the z-axis, where $\delta>0$. In this case, conical singularities, which act as stretched strings on the outer sides of the BHs, exert an outward pulling force. In this work, we restrict our analysis to asymptotically flat solutions.

\subsection{Ansatz and boundary conditions}

The focus of this section is to construct and analyze configurations describing 2sBHs with self-interacting scalar hair. To this end, we consider the corrections induced by the scalar field to the BW metric (\ref{BW_metric}), (\ref{weyl}) and generalize the metric into the spinning case:
\begin{eqnarray}\label{BW1}
ds^2= -e^{2(V+F_0)}dt^2 + 
e^{2(K-V+F_1)}(d\rho^2+dz^2)+
e^{2(-V+F_2)}\rho^2(d\varphi-Wdt)^2
\,,
\end{eqnarray}
where the metric functions $F_i~(i=0,1,2)$ and $W$ depend only on the coordinates $\rho$ and $z$. In the vacuum limit, these correction functions become trivial, and the BW metric is recovered.

However, directly constructing 2sBHs using the Weyl-type ansatz (\ref{BW1}) requires compactification of both coordinates $\rho$ and $z$, which leads to 
less accurate
ADM quantities extracted from the asymptotic behavior of the metric functions. Therefore, to improve numerical accuracy, we introduce the following coordinate transformation \cite{Herdeiro:2023mpt}
\begin{equation}\label{trans}
\rho = \frac{r^2-a^2}{r} \sin\theta, \quad z = \frac{r^2+a^2}{r} \cos\theta
\,,
\qquad
\text{with}
\qquad
m_h=\frac{(a-b)^2}{b}\,,
\qquad
\Delta z=4a\,,
\end{equation}
where $a \leq r < \infty$ and $0 \leq \theta \leq \pi$. The ansatz (\ref{BW1}) becomes
\begin{equation}
ds^2 = -f_0 e^{2F_0}dt^2+ f_1 e^{2F_1}(dr^2 + r^2 d\theta^2) + f_2 e^{2F_2}(dt - W d\phi)^2 
\,,
\end{equation}
where $f_i$ correspond to the vacuum BW solution. Under this transformation, the rod structure of 2sBHs consists of five segments: (i) two semi-infinite spacelike rods, $\theta=0$, $b< r < \infty$ and $\theta=\pi$, $b< r < \infty$; (ii) two timelike rods (event horizons), $\theta=0$, $a\leq r \leq b$ and $\theta=\pi$, $a\leq r \leq b$; (iii) a spacelike rod, $r=a$, $0\leq \theta \leq \pi$. 

After transforming the coordinates from $(\rho, z)$ to $(r, \theta)$, we still use the ansatz (\ref{matter_ansatz}) for the matter field. For 2sBHs, after substituting the above ansatz into (\ref{equek}), we can obtain a set of elliptic equations for numerically constructing 2sBHs. We impose the following boundary conditions to solve the equations of motion. At $r = a$,
\begin{equation}
\partial_r F_0|_{r=a}=\partial_r F_1|_{r=a}=\partial_r F_2|_{r=a}=\partial_rW|_{r=a}=0\,,\quad \psi|_{r=a}=0
\,.
\end{equation}
In addition, we impose the condition $F_1-F_0 = \text{const.}$ at $r = a$, i.e., $\delta$ is constant on rod (i). At $\theta = 0$ and $\pi$,
\begin{equation}
\partial_\theta F_0|_{\theta=0,\pi}=\partial_\theta F_1|_{\theta=0,\pi}=\partial_\theta F_2|_{\theta=0,\pi}=0\,,\quad W|_{\theta=0,\pi}=\Omega_H \,,\quad \partial_\theta\psi|_{\theta=0,\pi}=0
\,,
\end{equation}
where $\Omega_H = \omega/m$ is the horizon angular velocity of each BH, meaning that each BH possesses synchronised scalar hair. Additionally, for $a \leq r \leq b$, i.e., rods (ii), we impose $F_0-F_1 = \text{const.}$, which means the Hawking temperature is constant. For $r > b$, we impose the constraint $F_1 = F_2$ to avoid conical singularities on rod (iii). Asymptotic flatness requires that all functions vanish at infinity,
\begin{equation}
F_0|_{r\to \infty}=F_1|_{r\to \infty}=F_2|_{r\to \infty}=W|_{r\to \infty}=\psi|_{r\to \infty}=0
\,,
\end{equation}
Moreover, similar to the case of 1sBHs, the scalar field amplitude changes sign across the equatorial plane, while the geometry is $\mathbb{Z}_2$-even. Therefore, at $\theta = \pi/2$,
\begin{equation}
\partial_\theta F_0|_{\theta=\frac{\pi}{2}}=\partial_\theta F_1|_{\theta=\frac{\pi}{2}}=\partial_\theta F_2|_{\theta=\frac{\pi}{2}}=\partial_\theta W|_{\theta=\frac{\pi}{2}}=0
\,, \quad \psi|_{\theta=\frac{\pi}{2}}=0
\,.
\end{equation}
For 2sBHs, we also calculate the physical quantities mentioned in Subsection \ref{quantities}. In addition, we define the distance between the two event horizons as
\begin{equation}\label{distance}
L = \int_{0}^{\pi}  \,d\theta \sqrt{g_{\theta\theta}(a,\theta)} .
\end{equation}
In this way, a given set of $(\omega, \lambda, M, L)$ uniquely determines a 2sBH solution.

\begin{figure}[h!]
  \centering
  \includegraphics[width=0.49\textwidth]{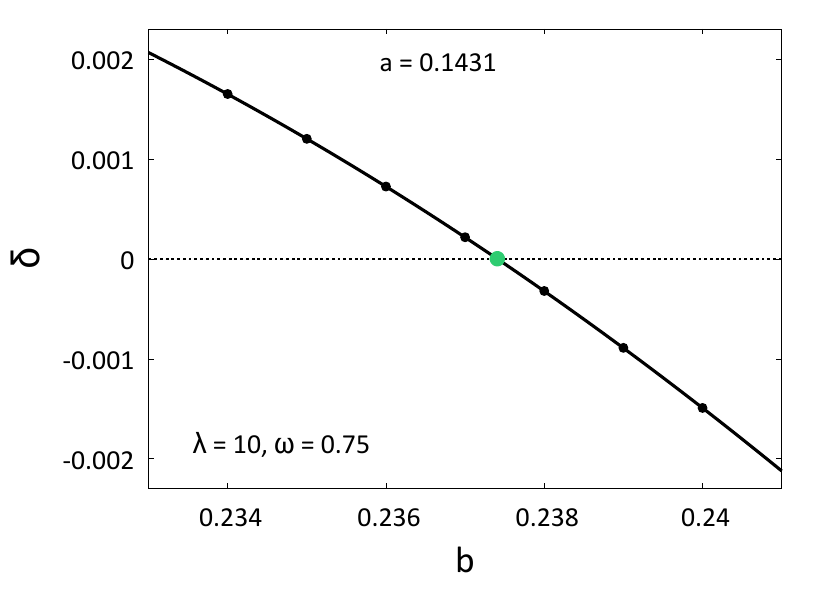}
  \includegraphics[width=0.49\textwidth]{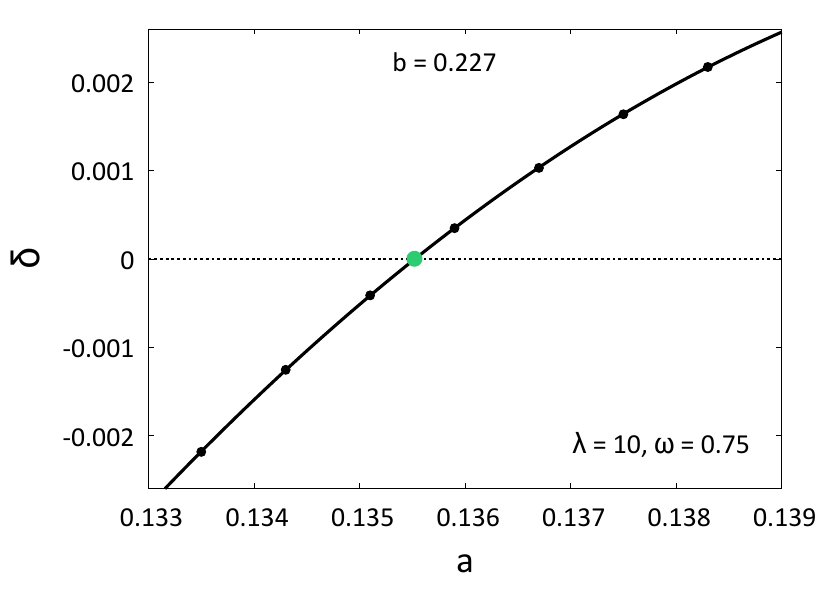}
  \\[20pt]
  \includegraphics[width=0.6\textwidth]{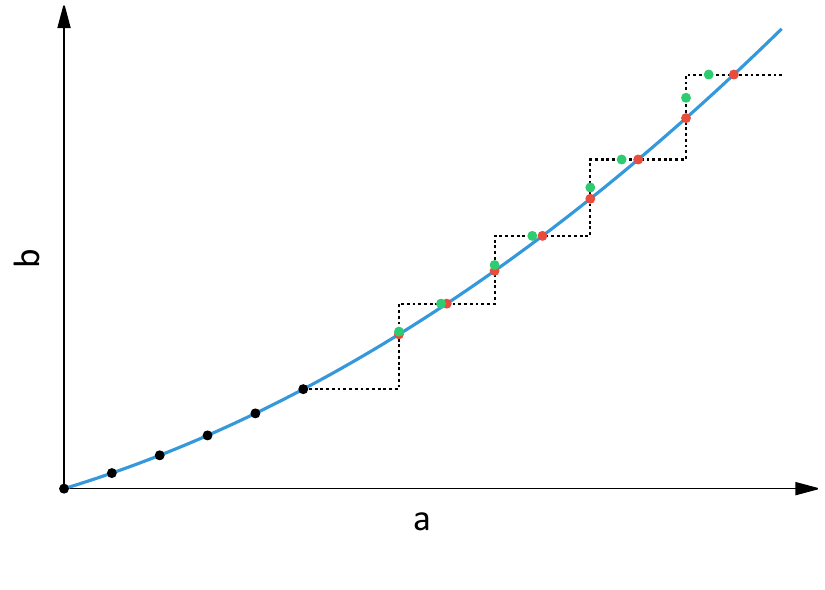}
  \caption{(Top panels) The conical excess/deficit $\delta$ as a function of the parameters $(a, b)$. (Bottom panel) A curve (blue line) fitted in the parameter space $(a, b)$ using several balanced solutions (black points). The parameters $(a, b)$ are alternately scanned along this curve, with the scanning path forming a staircase pattern (black dashed line).}
  \label{path}
\end{figure}
\begin{figure}[h!]
  \centering
  \includegraphics[width=0.48\textwidth]{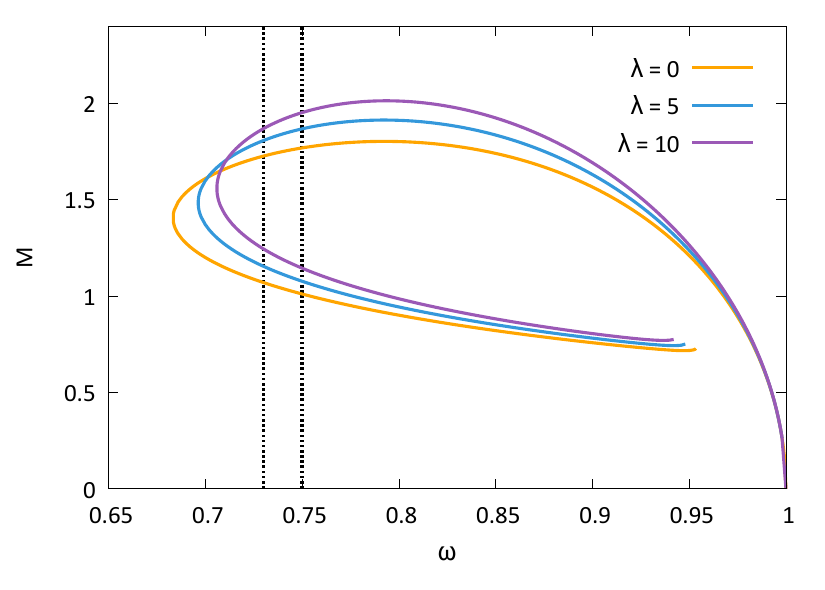}
  \hspace{0.05\textwidth}
  \includegraphics[width=0.4\textwidth]{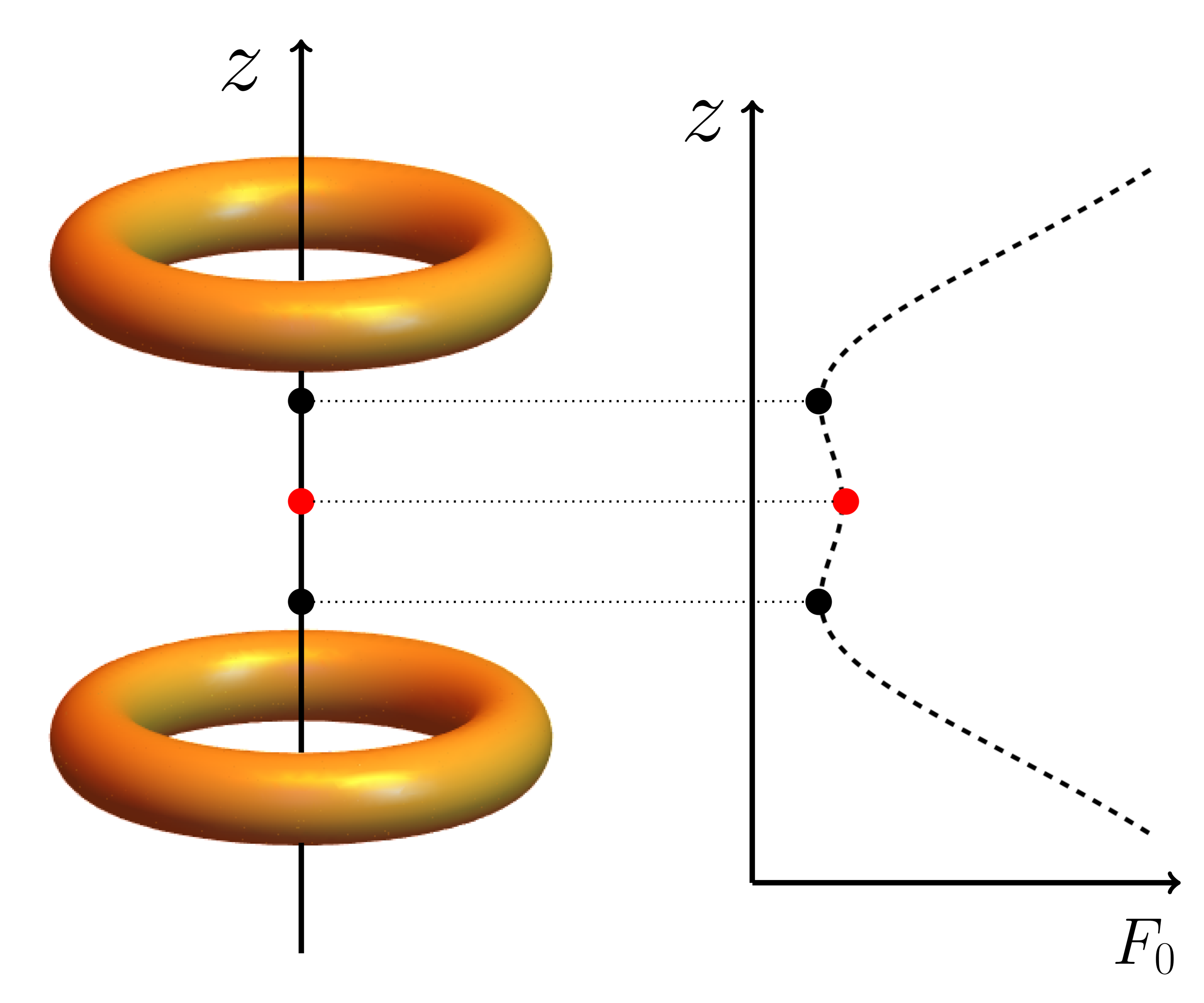}
  \caption{(Left) ADM mass $M$ \textit{vs.} frequency $\omega$ for 2sBSs with $\lambda=0, 5, 10$. The vertical dashed lines correspond to $\omega=0.73, 0.75$. (Right) The 2sBS as a scalar environment for constructing 2sBHs. Red and black dots mark the unstable and stable equilibrium points along the symmetry axis, respectively.} 
  \label{Mw}
\end{figure}

\subsection{Numerical results}
The 2sBHs we construct are in equilibrium, where the $\pi$ phase difference of the scalar cloud provides an additional repulsive force to balance gravity, so that the configuration does not require a conical singularity for support. However, the conical excess/deficit $\delta$ is highly correlated with the parameters $a$ and $b$ in (\ref{trans}),
\begin{equation}
\delta = 2\pi\left(1-\frac{(a+b)^4e^{F_2(a,\theta)-F_1(a,\theta)}}{8ab(a^2+b^2)}\right),
\end{equation}
which requires us to adopt an appropriate strategy to construct these equilibrium solutions. As shown in the upper panels of Fig. \ref{path}, the black points represent non-equilibrium solutions. For a given set of $(\lambda, \omega)$, $\delta$ varies monotonically with $b$ ($a$) when $a$ ($b$) is held fixed, and the monotonicity of $\delta$ with respect to $a$ and $b$ is opposite. By fitting a curve through the black points, i.e., the black solid 
line in the figure, we can estimate the values of $a$ and $b$ corresponding to $\delta = 0$, thereby constructing an equilibrium solution, marked as the green point.\footnote{In practice, we can only make $\delta$ as close to zero as possible. For the solutions presented in this work, $\delta \sim 10^{-6}$.} To obtain a complete solution branch of 2sBHs, our strategy is roughly as follows:
\begin{enumerate}[label=(\alph*)]
  \item Using the above method, we obtain several equilibrium solutions $\{(a_1, b_1), (a_2, b_2), \ldots, (a_n, b_n)\}$, as shown by the black points in the bottom panel of Fig. \ref{path}.
  \item We fit these $n$ points in the parameter space $(a, b)$, yielding the blue curve in the plot. Along this curve, we alternately scan the parameters $a$ and $b$. The staircase-like dashed line represents the scanning path. The intersections of the path with the blue line are taken as the predicted values of $(a, b)$ corresponding to equilibrium solutions, marked as red points.
  \item We interpolate along each segment of the path to obtain equilibrium solutions, i.e., the green points in the plot.
  \item By repeating the steps (b) and (c), we can obtain a complete solution branch.
  These solutions do not possess a conical singularity
  or any other pathology, on or outside the event horizons.
\end{enumerate}

In this work, we consider the cases of $\omega=0.73$ and $0.75$, as shown in the left panel of Fig.~\ref{Mw}. Following the strategy outlined above, we construct 2sBHs with $\lambda=0$, $5$, and $10$. For fixed $\omega$ and 
$\lambda$, we introduce the proper distance~(\ref{distance}) to display the domain of existence of 2sBHs, which consists of a series of solution sequences bifurcating from either the upper or lower branches of 2sBSs.

\begin{figure}[h!]
  \centering
  \includegraphics[width=0.49\textwidth]{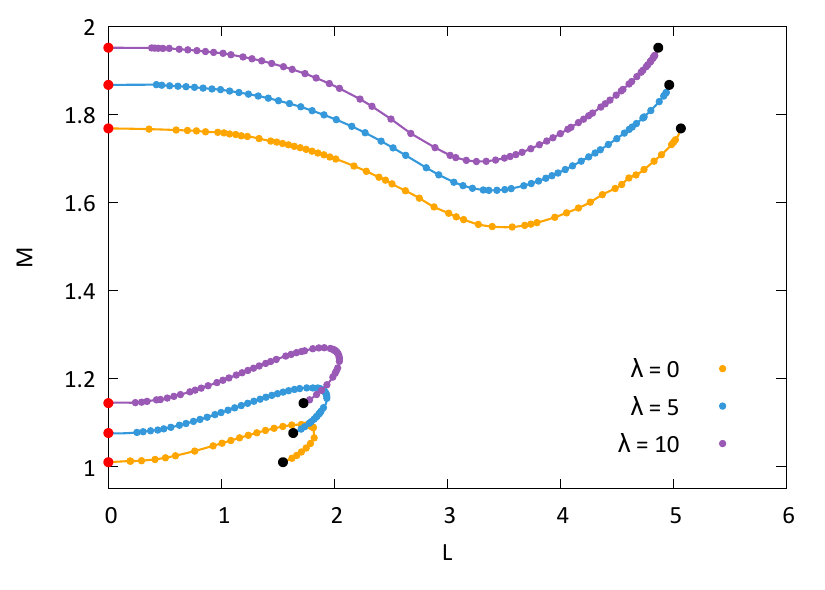}
  \includegraphics[width=0.49\textwidth]{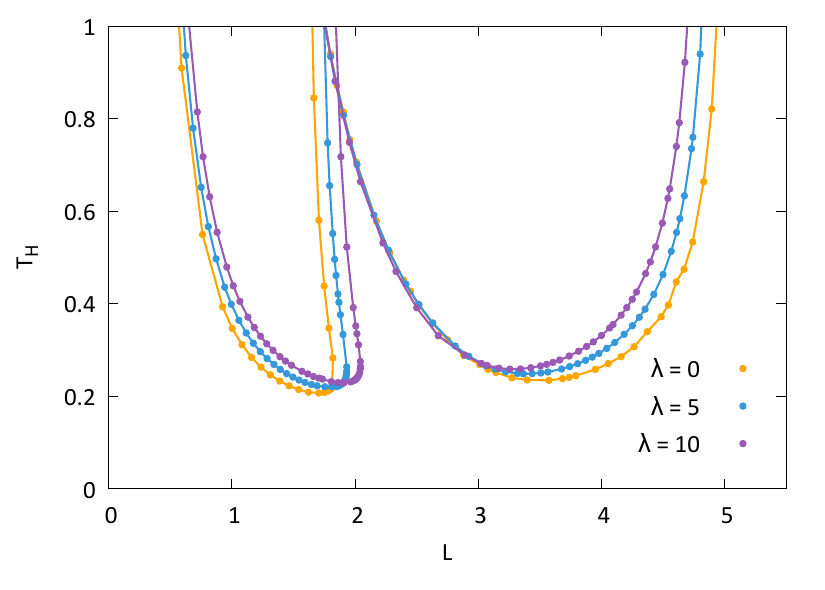}
  \includegraphics[width=0.49\textwidth]{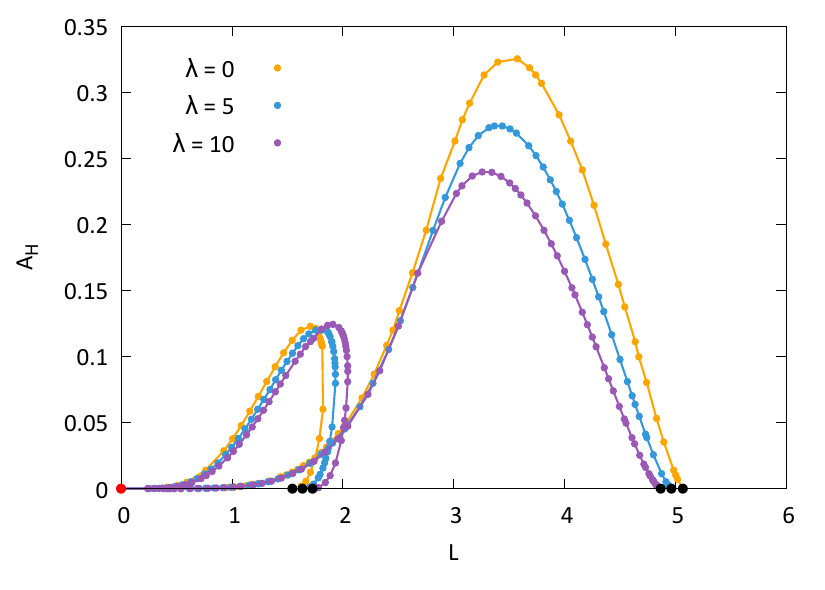}
  \includegraphics[width=0.49\textwidth]{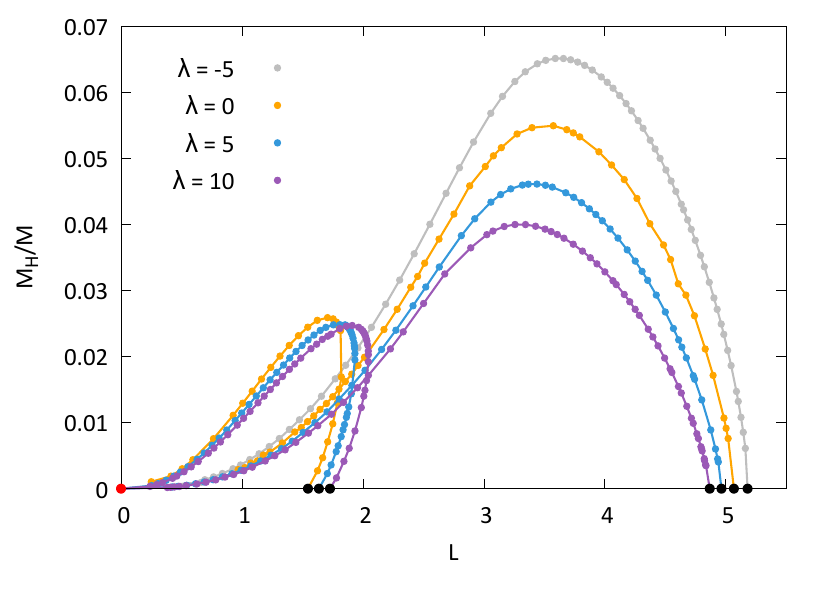}
  \caption{ADM mass $M$ (top left panel), Hawking temperature $T_H$ (top right panel), horizon area $A_H$ (bottom left panel) and ratio $M_H/M$ \textit{vs.} distance $L$ for 2sBHs with $\omega=0.75$ and $\lambda=0, 5, 10$. }
  \label{w0.75}
\end{figure}

To facilitate the understanding of the 2sBHs constructed here, the right panel of Fig.~\ref{Mw} provides a qualitative illustration of the relationship between 2sBSs and 2sBHs. The double-torus surface on the left depicts the matter distribution of 2sBSs, while the curve on the right shows the profile of the metric function $F_0$ in the ansatz~(\ref{metric}) along the $z$-axis, which passes through the centers of the two tori. In Newtonian gravity, consider a pair of massive rings placed along the $z$-axis, with the axis passing through both centers and perpendicular to the planes of the rings. When the separation between the rings is sufficiently large, the gravitational potential of the system can possess two local minima and one local maximum, corresponding to two stable equilibrium points and one unstable equilibrium point, respectively. For 2sBSs, this double-torus matter distribution endows the gravitational potential with precisely these characteristics. As shown in the right panel, the metric function $F_0$ always exhibits three extrema on the $z$-axis. At these extrema, a massive test particle can remain stationary, satisfying the equilibrium conditions
\begin{equation}
    (\dot{\rho}, \dot{z}, \dot{\varphi}) = (\ddot{\rho}, \ddot{z}, \ddot{\varphi})= 0\,,
\end{equation}
together with the geodesic equation
\begin{equation}
    \ddot{x}^{\mu} + \Gamma^{\mu}_{\nu \gamma}\dot{x}^{\nu}\dot{x}^{\gamma} = 0\,,
\end{equation}
where the overdots denote derivatives with respect to proper time. Considering the metric ansatz~\eqref{metric}, the geodesic equation yields $\partial_z F_0 = 0$ at the equilibrium points~\cite{Herdeiro:2023roz}. Thus, one local maximum and two local minima of $F_0$ on the $z$-axis correspond to one unstable and two stable equilibrium points, respectively. A 2sBS can thus be viewed as a scalar environment in which the two horizons may either both grow from the unstable equilibrium point or grow separately from the two stable equilibrium points. In other 
words, each sequence of 2sBHs possesses two bifurcation points at which it intersects 2sBSs, and at these bifurcation points the two event horizons either both emerge from the unstable point or emerge separately from the 
stable points.

According to the nature of the endpoints, these sequences can be classified into the following three types:
\begin{itemize}
    \item[(i)] sequences with horizons emerging from the stable points at one endpoint and from the unstable point at the other;
    \item[(ii)] sequences with horizons emerging from the unstable point at both endpoints;
    \item[(iii)] sequences with horizons emerging from the stable points at both endpoints.
\end{itemize}

\begin{figure}[h!]
  \centering
  \includegraphics[width=0.49\textwidth]{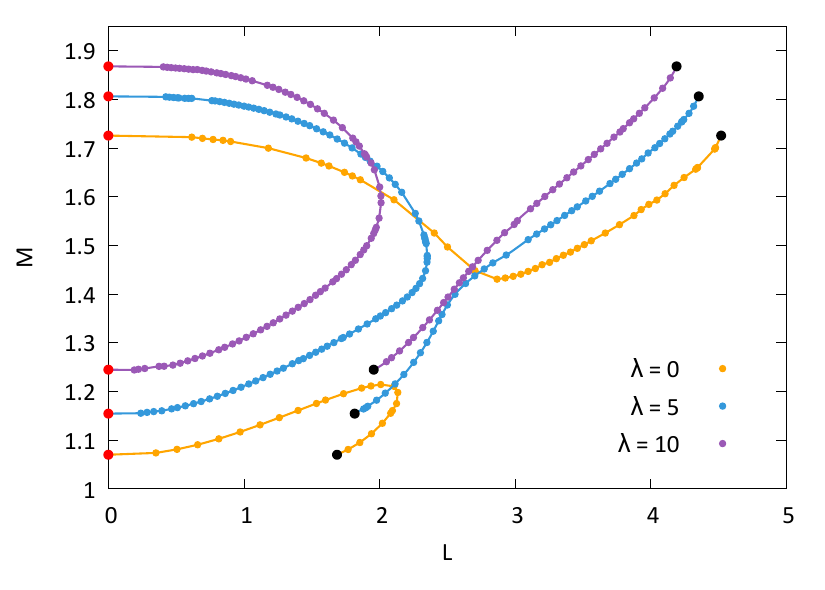}
  \includegraphics[width=0.49\textwidth]{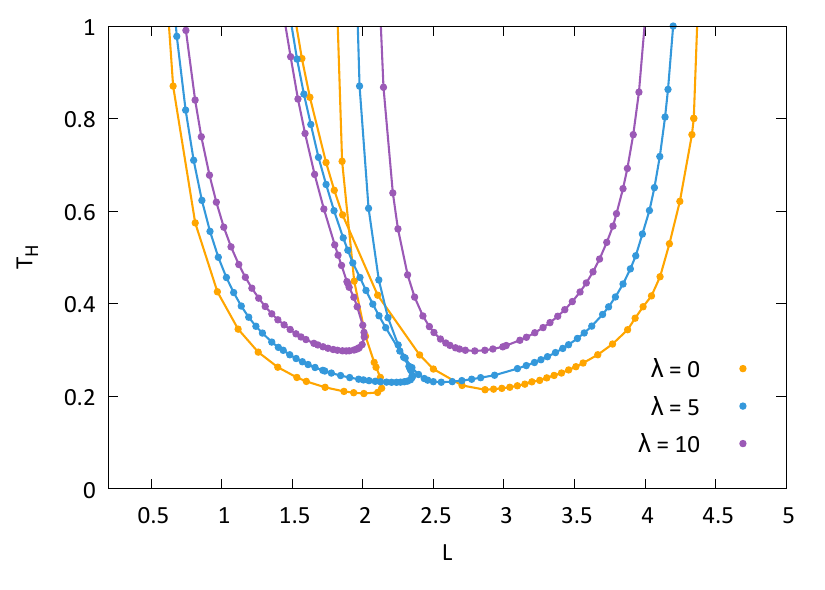}
  \includegraphics[width=0.49\textwidth]{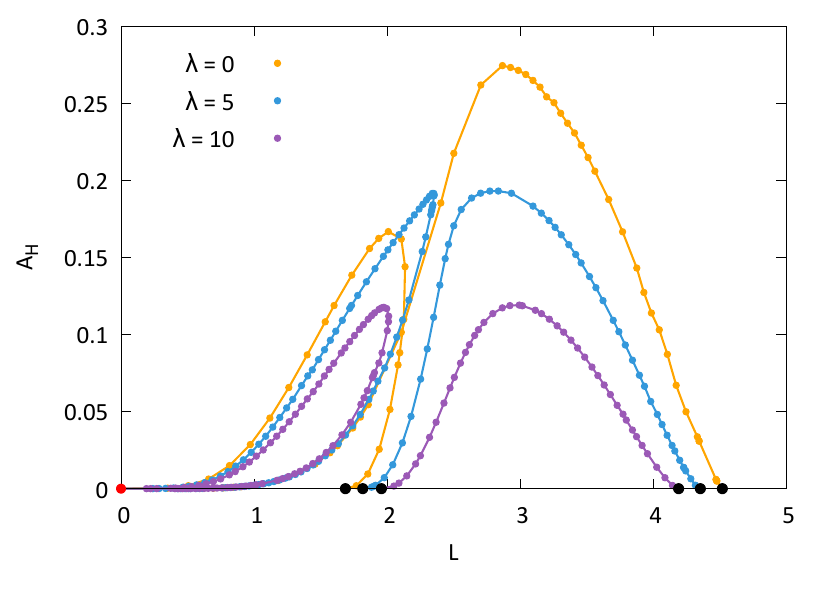}
  \includegraphics[width=0.49\textwidth]{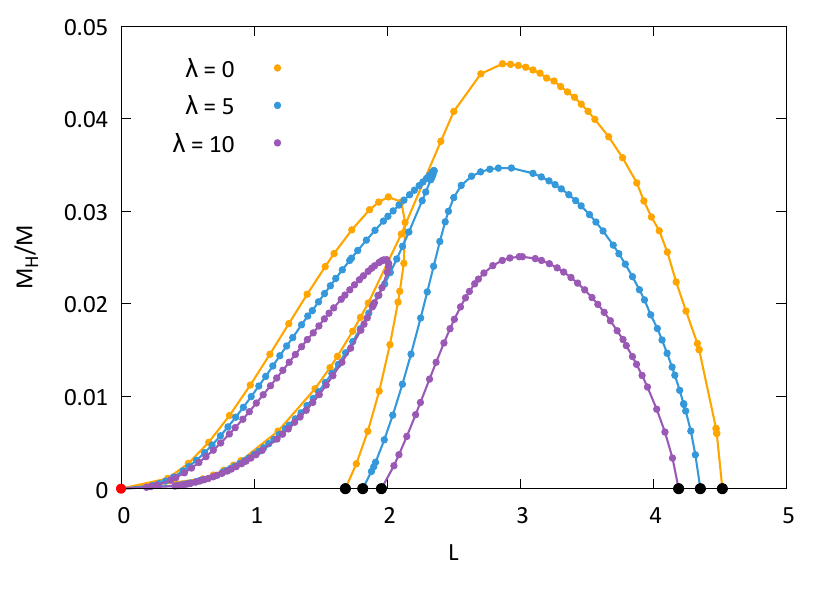}
  \caption{Same as Fig. \ref{w0.75} for 2sBSs with $\omega=0.73$.} 
  \label{w0.73}
\end{figure}

Among these, sequences~(i) always appear in pairs, as shown in Fig.~\ref{w0.75}. The top left panel displays the relation between the mass and the proper distance between the two horizons. For a given $\lambda$, the upper and lower sequences~(i) bifurcate from 2sBSs lying on the upper and lower branches of the 2sBS solutions, respectively. 
The values of $L$ along the upper sequences~(i) always remain between those at the two endpoints, indicating that each horizon of these 2sBHs is confined between a stable and the unstable equilibrium point along the $z$-axis. The lower sequences~(i) behave differently: the horizons of the corresponding 2sBHs can be located slightly beyond the stable equilibrium points along the $z$-axis. As in the case of 1sBHs, the mass of sequences~(i) is bounded by that of 2sBSs, and consequently the mass increases with $\lambda$. A mass gap is clearly visible between the upper and lower sequences of each pair. By analogy with the 1sBH case, this mass gap may be associated with extremal 2sBHs, although their existence remains uncertain at present.

The top right panel shows that the minimum of Hawking temperature $T_H$ increases as the coupling constant is increased. As $L$ and $\lambda$ vary, $T_H$ of the 2sBHs cannot be arbitrarily small. This does not, however, rule out the existence of extremal 2sBHs, since the equilibrium condition $\delta=0$ may impose a constraint such that extremal and non-extremal 2sBHs are connected through 2sBHs with $\delta\neq 0$.

Sequences~(ii) and~(iii) always appear together. As shown in Fig.~\ref{w0.73}, as $\lambda$ increases, the mass gap between the upper and lower sequences~(i) closes, and the two sequences transform into one sequence~(ii) and one sequence~(iii), with the mass of the 2sBHs still bounded by that of 2sBSs. Interestingly, for given $\omega$ and $\lambda$, sequences~(ii) and~(iii) exhibit unexpectedly close extremal values in the horizon area, the Hawking temperature, and the ratio $M_H/M$.

The lower panels of Figs.~\ref{w0.75} and~\ref{w0.73} show that, as $\lambda$ increases, the strengthened repulsive self-interactions suppress the size of the horizons and accordingly reduce the fraction of the total mass contained in the horizons, making the horizons lighter. By contrast, if a negative (attractive) coupling constant is considered---for which the scalar potential becomes unbounded from below---the horizons of the 2sBHs can carry a larger mass fraction than in the non-self-interacting case, becoming heavier, as shown in the lower right panel of Fig.~\ref{w0.75}. We further note that the ratio $M_H/M$ always remains at the level of a few percent, which is why the 2sBHs constructed here can be qualitatively described using the scalar environment illustrated in the right panel of Fig.~\ref{Mw}.

\section{Conclusions}\label{conclusions}

In this work, we have investigated a family of asymptotically flat, stationary configurations of Einstein gravity minimally coupled to a complex scalar field with a repulsive quartic self-interaction. The family consists of three related models sharing the same two-lump structure of the scalar field: the 2sBSs, the 1sBHs obtained by placing a horizon at the center of a 2sBS, and the 2sBHs obtained by placing two aligned horizons along the symmetry axis of a 2sBS. For the 2sBHs, we generalized the Bach-Weyl framework to the spinning, hairy 
self-interacting case, and developed a 
numerical strategy based on alternate scanning in the parameter space $(a,b)$ to construct equilibrium solutions with vanishing conical singularity.

\medskip

Our main findings can be summarized as follows. For 2sBSs, the quartic self-interaction shifts the domain of existence towards larger ADM masses. Ergoregions are found to arise only in the strong-gravity regime, where the two components are sufficiently close; as $\lambda$ increases, the ergoregion undergoes a topological transition from a single torus to a double torus, driven by the self-interaction-induced redistribution of scalar matter away from the equatorial region. For 1sBHs, the domain of existence is bounded by 2sBSs, the Kerr existence line, and the extremal line, and always contains a subset of solutions violating the Kerr bound. The horizon quantities $(M_H,J_H)$ remain bounded by the Hod point independently of $\lambda$, showing that the ``hairier but not heavier'' behavior previously found for the fundamental scalar hair~\cite{Herdeiro:2015tia} extends to the quadrupolar case. We also find that self-interactions enlarge the range of $p$ in which the effective model of~\cite{Herdeiro:2017phl, Brihaye:2018woc} provides an accurate description of 1sBHs. For 2sBHs, the solution sequences are classified into three types according to how their horizons emerge from the equilibrium points of the scalar environment. As $\lambda$ increases, the mass gap between the two sequences of type (i) closes and the pair transforms into one sequence of type (ii) and one of type (iii), whose extremal values of the horizon area, the 
Hawking temperature, and the ratio $M_H/M$ are remarkably close. As in the 1sBH case, repulsive self-interactions cannot make the horizons heavier; horizons carrying a larger mass fraction are obtained only when attractive self-interactions are considered.

Several directions remain open for future investigation. A natural question is how the present results compare with other balanced multi-black-hole configurations. In the electrovacuum case, the Majumdar-Papapetrou solution describes multiple extremal BHs in static equilibrium, where the mutual gravitational attraction is balanced by electrostatic repulsion~\cite{Majumdar:1947eu, Papapetrou:1948jw}. Rotating generalizations of this configuration are known in Einstein-Maxwell-dilaton theory~\cite{Teo:2023wfd,Moreira:2025ckc}, but not  in pure Einstein-Maxwell gravity and flat spacetime asymptotics. 
It would therefore be interesting to investigate 
how additional fields---such as 
a self-interacting scalar---can create balanced charged, spinning double BH solutions. 
Another promising direction is to consider Proca hair as a substitute for the scalar field. Following the test-particle picture developed in Sec.~\ref{sec:2sBHs}, a necessary step in this direction would be the construction of spinning Proca stars admitting two stable equilibrium points along the symmetry axis, analogous to the 2sBSs studied here; however, such configurations have not yet been reported in the literature. The corresponding hairy BHs, if they exist, would provide a vector-field counterpart to the 2sBHs and offer further insight into the role of the matter content in supporting equilibrium configurations of multiple rotating horizons.

\section*{Acknowledgments}
C.L. thanks E. dos Santos Costa Filho for useful discussions. This work is supported by the Center for Research and Development in Mathematics and Applications (CIDMA) (\url{https://ror.org/05pm2mw36}) under the Portuguese Foundation for Science and Technology 
(FCT -- Fundaç\~ao para a Ci\^encia e a Tecnologia, \url{https://ror.org/00snfqn58}), Grants UID/04106/2025 (\url{https://doi.org/10.54499/UID/04106/2025}) and UID/PRR/04106/2025 (\url{https://doi.org/10.54499/UID/PRR/04106/2025}), as well as the projects: Horizon Europe staff exchange (SE) programme HORIZON-MSCA2021-SE-01 Grant No.\ NewFunFiCO-101086251 and 2022.04560.PTDC (\url{https://doi.org/10.54499/2022.04560.PTDC}).
C.L. \ is supported by China Scholarship Council (CSC) fellowship.
The authors thankfully acknowledge computational resources from the Argus and Blafis clusters at the University of Aveiro.

\appendix

\section{The vacuum Bach-Weyl solution in $(r,\theta)$ coordinates}
\label{appA}
Under the coordinate transformation (\ref{trans}), the Bach-Weyl solution (\ref{weyl}) can be transformed into the following form
\begin{subequations}\label{weylnew}
\begin{align}
f_0(r,\theta)
&=
\frac{f^{(1)}_{0+}(r,\theta) f^{(1)}_{0-}(r,\theta)}{f^{(2)}_{0+}(r,\theta) f^{(2)}_{0-}(r,\theta)}
\,,
\\
f_1(r,\theta)
&=
\frac{f^{(1)2}_1(r,\theta)}{f_0(r,\theta) f^{(2)}_1(r,\theta)} \frac{f^{(3)}_{1+}(r,\theta)f^{(3)}_{1-}(r,\theta)}{f^{(4)2}_{1+}(r,\theta)}
\,,
\\
f_2(r,\theta)
&=
\frac{(a^2-r^2)\sin^2\theta}{r^2f_0(r,\theta)}
\,,
\end{align}
\end{subequations}
with
\begin{subequations}
\begin{align}
f^{(1)}_{0\pm }(r,\theta)
&=
(b - r)(a^2 - br) + 2abr \pm 2abr\cos \theta + A_\pm B_\pm
\,,
\\
f^{(2)}_{0\pm }(r,\theta)
&=
(b + r)(a^2 + br) - 2abr \pm 2abr\cos \theta + A_\pm B_\pm
\,,
\\
f^{(1)}_1(r,\theta)
&=
2 b \left(a^2+b^2\right) \left(a^2+r^2\right)-4 a b r \left(a^2+b^2\right) \cos \theta +
a^2(A_-B_- - A_+B_+) 
\notag
\\
&+ 
b^2(A_-B_- - A_+B_+) + 2ab(A_-B_- + A_+B_+)
\,,
\\
f^{(2)}_1(r,\theta)
&=
4r^4 (a+b)^4(A_+A_-B_+B_-)
\,,
\\
f^{(3)}_{1\pm}(r,\theta)
&=
a^4 b+2 a r^2 \left(a^2+b^2\right)+2 a^2 b r^2 \cos 2 \theta \pm r (a+b)^2 \left(a^2+r^2\right) \cos \theta +b r^4
\notag
\\
&+
 (a^2+r^2 \pm 2ar\cos \theta)A_\pm B_\pm
 \,,
\\
f^{(4)}_1(r,\theta)
&=
b(a^2+r^2)-(a^2+b^2)r\cos \theta +A_-B_-
\,
\end{align}
\end{subequations}
and
\begin{align}
A_\pm = \sqrt{a^4+b^2r^2 \pm 2a^2br\cos \theta}\,, \quad B_\pm = \sqrt{b^2 + r^2 \pm 2br\cos \theta}
\,.
\end{align}
Therefore, in this coordinate system, the physical quantities~(\ref{mj}) and~(\ref{thah}) are expressed as
\begin{subequations}
  \begin{align}
\notag
M_H & =-\frac{1}{8\pi G}\oint_{\mathcal{H}}dS_{\mu\nu}\nabla^\mu\xi^\nu \\
&= 
\frac{1}{G}\int_{a}^{b} dr \left(\frac{16e^{3F_2-F_0}(r-a)^3(b^2-r^2)^2(a^4-b^2r^2)^2\omega W_{,\theta\theta}}{b^4 m r^4(r+a)^5} - \frac{e^{F_0+F_2} (r^2-a^2) }{r^2}\right)\Big|_{\theta=0}
\,,
\\\notag
J_H & =\frac{1}{16\pi G}\oint_{\mathcal{H}}dS_{\mu\nu}\nabla^\mu\eta^\nu \\
&=
-\frac{1}{G}\int_{a}^{b} dr \frac{8e^{-F_0+3F_2}(a-r)^3(b^2-r^2)^2(a^4-b^2r^2)^2 W_{,\theta\theta}}{b^4r^4(r+a)^5}\Big|_{\theta=0}
\,,\\\notag
\\\notag
M_\Psi & =-\frac{1}{4\pi G}\int_\Sigma dS_\mu(2T_\nu^\mu\xi^\nu-T\xi^\mu)\\
&= \frac{1}{G}\int_{a}^{\infty} dr \int_{0}^{\pi} d\theta e^{F_0+2F_1+3F_2}r \sqrt{f_0f_1^2f_2} \left( U - \frac{2e^{-2F_0} \omega (\omega-mW)}{f_0} \psi^2 \right)
\,,\\\notag
\\\notag
J_\Psi & =\frac{1}{8\pi G}\int_\Sigma dS_\mu\left(T_\nu^\mu\eta^\nu-\frac{1}{2}T\eta^\mu\right)\\
&= \frac{1}{2G}\int_{a}^{\infty} dr \int_{0}^{\pi} d\theta \frac{e^{-F_0+2F_1+3F_2}m r \sqrt{f_0f_1^2f_2}\psi^2}{f_0}(\omega-mW)\,,
\end{align}
\end{subequations}
and
\begin{align}
T_H=\frac{b(a+b)^2}{8(a-b)^2(a^2+b^2)\pi}e^{F_0-F_1-F_2}\Big|_{\theta=0}
\,,\quad
A_H=2\pi \int_{a}^{b} dr \frac{4(a-b)^2(a^2+b^2) e^{F_1+2F_2}(r^2-a^2)}{b(a+b)^2r^2}\Big|_{\theta=0}
\,.
\end{align}
Here, $T_H$ and $A_H$ are constant on the horizon, i.e., for $r\in[a,b]$, which also provides a check on the accuracy of our numerical method by verifying that these two quantities are uniform along the horizon.

\bibliographystyle{unsrt}
\bibliography{refs}

\end{document}